\PassOptionsToPackage{
	colorlinks=true,
	citecolor=blue,
	linkcolor=red,
	urlcolor=blue,
	anchorcolor=red,
	menucolor=red,
	filecolor=red,
	runcolor=red
}{hyperref}

\documentclass[aps,prd,floatfix,onecolumn,amsmath,amssymb,showpacs,showkeys]{revtex4-2}

\usepackage{graphicx}
\usepackage{xcolor}
\usepackage{dcolumn}
\usepackage{booktabs}
\usepackage{bm}
\usepackage{multirow}

\usepackage{slashed}
\usepackage{cases}
\usepackage{pgf}

\usepackage{hyperref}

\usepackage{subcaption}
\definecolor{maroon}{RGB}{139,25,150}

\begin{document}
	
	\preprint{}
	
	\title{Numerical Study of MRW-Type Unintegrated Double Parton Distribution Functions from Non-Factorized DPDFs}
	
	\author{The CHROMA Collaboration:\\
		R.~Kord Valeshabadi$^{1}$}
	
	\email[Corresponding author: ]{ramin.kord@ipm.ir}
	
	\author{S.~Rezaie$^{1}$}
	\email{somayeh@ipm.ir}
	
	\author{K.~Azizi$^{2,3,1}$}
	\email{kazem.azizi@ut.ac.ir}

	\affiliation{
		$^{1}$School of Particles and Accelerators, Institute for Research in Fundamental Sciences (IPM), P.O. Box 19568-36681, Tehran, Iran\\
		$^{2}$Department of Physics, University of Tehran, North Karegar Avenue, Tehran 14395-547, Iran\\
		$^{3}$Department of Physics, Dogus University, Dudullu-Umraniye, 34775
		Istanbul, Turkiye
	}
	
	\begin{abstract}
	Double parton scattering provides a sensitive probe of multiparton correlations inside hadrons. In this work we present a numerical study of unintegrated double parton distribution functions constructed from non-factorized collinear DPDFs. As input we use the \texttt{GS09} DPDFs and evolve them to unequal scales with a numerical double DGLAP evolution framework, which is validated through the corresponding momentum and valence-number sum rules.
		
	We investigate MRW-inspired prescriptions for generating transverse momentum dependence in
	the double parton case. In particular, we study the double LO-MRW approach
	(DLO-MRW) in the perturbative region, the double modified KMRW approach (DMKMRW),
	the double virtuality ordered MRW (DVO-MRW) model, and a normalization-matched version
	of the latter. The DMKMRW, DVO-MRW, and matched DVO-MRW prescriptions can be applied
	directly to the full collinear DPDF, including its nonhomogeneous component, and avoid the
	piecewise treatment required in the conventional LO-MRW construction.
	
	We analyze the normalization, transverse momentum dependence, flavor dependence, and sensitivity to longitudinal DPDF correlations of the resulting distributions. The DMKMRW model is normalization preserving by construction, while the DVO-MRW model shows nontrivial normalization deviations that are removed in the matched version. Non-factorized effects are strongly channel-dependent. In DMKMRW, they enter mainly through longitudinal DPDF correlations, whereas in the LO-MRW and virtuality-ordered models, they also modify the transverse momentum shape. The largest deviations occur near the double parton kinematic boundary and in channels affected by valence-number and quark-antiquark correlations.
	\end{abstract}
	
	
	\maketitle
	\newpage
	\section{Introduction}
	
	Hadron-hadron scattering at high-energy colliders is usually described within the framework of QCD factorization. In this approach, hadronic cross sections are written as convolutions of partonic hard scattering cross sections with parton distribution functions (PDFs). In the simplest picture, a single parton from one hadron interacts with a single parton from the other hadron. This mechanism, known as single parton scattering (SPS), provides the dominant contribution to many hard processes.

	At sufficiently high center of mass energies, however, the probability that more than one pair of partons undergoes hard interactions in the same hadron-hadron collision becomes non-negligible. The simplest such contribution is double parton scattering (DPS), in which two separate hard subprocesses occur in a single collision. Experimental evidence for DPS was first reported by the AFS Collaboration at CERN~\cite{AxialFieldSpectrometer:1986dfj}. Among more recent measurements, the CMS study of low transverse momentum four-jet production at $\sqrt{s}=13~\mathrm{TeV}$~\cite{CMS:2021lxi} is particularly relevant for the present work, since four-jet DPS observables can be sensitive to the transverse momenta of the incoming partons.
	
	The DPS cross section can be expressed in terms of double parton distribution functions (DPDFs) and two partonic cross sections. The generalized DPDFs, $\Gamma_{i,j}(x_1,x_2;b;\mu_A^2,\mu_B^2)$,
	describe the probability of finding two partons of flavors $i$ and $j$ in the proton, carrying longitudinal momentum fractions $x_1$ and $x_2$, probed at the scales $\mu_A^2$ and $\mu_B^2$, and separated by a transverse distance $b$. The theoretical structure of DPDFs and DPS has been studied extensively; see, for example, Refs.~\cite{Diehl:2021wvd,Diehl:2021wpp,Diehl:2018wfy,Buffing:2017mqm,Diehl:2017kgu,Diehl:2017wew,Diehl:2020xyg}. 
	
	A commonly used phenomenological assumption is that the $b$ transverse dependence of generalized form of DPDFs can be factorized from the longitudinal one,
	\begin{equation}
		\label{eq:b_factorized_ansatz}
		\Gamma_{i,j}(x_1,x_2,b,\mu_A^2,\mu_B^2)
		=
		D_{i,j}(x_1,x_2,\mu_A^2,\mu_B^2) F_{i,j}(b),
	\end{equation}
	where $D_{i,j}$ denotes the collinear double parton distribution function. In addition, in many phenomenological applications the longitudinal DPDF is approximated by a product of two ordinary PDFs,
	\begin{equation}
		D_{i,j}(x_1,x_2,\mu_A^2,\mu_B^2)
		\simeq
		f_i(x_1,\mu_A^2) f_j(x_2,\mu_B^2)\Theta(1-x_1-x_2).
	\end{equation}
	Together with the assumption that $F_{i,j}(b)$ is the same for all parton pairs, i.e. $F_{i,j}(b) = F(b)$, this leads to the well-known pocket formula,
	\begin{equation}
		\sigma^{A,B}_{\mathrm{DPS}}
		=
		\frac{
			\sigma^{\mathrm{SPS}}_A
			\sigma^{\mathrm{SPS}}_B
		}{
			(1+\delta_{AB})\sigma_{\mathrm{eff}}
		},
	\end{equation}
	where $\delta_{AB}$ accounts for identical final states and
	\begin{equation}
		\sigma_{\mathrm{eff}}
		=
		\left[
		\int d^2 b \, \big(F(b)\big)^2
		\right]^{-1}.
	\end{equation}
	
	Although the pocket formula is useful for phenomenological estimates, it relies on strong simplifying assumptions. In particular, the factorized form of the collinear DPDFs should be regarded only as an approximation. Ordinary single PDFs obey the standard DGLAP evolution equations~\cite{DGLAP1,DGLAP2,DGLAP3}, whereas DPDFs satisfy the double DGLAP evolution equations~\cite{Ceccopieri:2010kg,Gaunt:2009re}. These equations describe the simultaneous scale dependence of two partons participating in two hard subprocesses and generate nontrivial longitudinal correlations between them. Therefore, the simple product ansatz in terms of two single parton distribution functions (SPDFs) does not, in general, provide a consistent solution of the double DGLAP equations. It also does not satisfy the DPDF momentum and number sum rules in general, even in the small-$x$ region~\cite{Gaunt:2009re,Golec-Biernat:2022wkx}. A more complete description of DPS therefore requires longitudinal correlations generated both by the nonperturbative proton structure and by perturbative QCD evolution.
	
	Another limitation of the standard collinear DPS framework is that the transverse momenta of the incoming partons are integrated out. At high energies, initial state QCD radiation can generate sizeable transverse momenta, which may affect the kinematic distributions of DPS final states. This motivates the introduction of unintegrated double parton distribution functions (UDPDFs), which retain the dependence on the transverse momenta of the two active partons. In analogy with unintegrated PDFs (UPDFs) in SPS, UDPDFs provide a more differential description of DPS and allow one to study DPS processes within the $k_t$-factorization framework~\cite{kt_factorization1,kt_factorization2}. This framework is closely related to high-energy QCD evolution approaches such as the Balitsky-Fadin-Kuraev-Lipatov (BFKL)~\cite{BFKL1,BFKL2} and Catani-Ciafaloni-Fiorani-Marchesini (CCFM)~\cite{CCFM1,CCFM2,CCFM3,CCFM4} equations. However, these aforementioned evolution equations are limited to the gluon case. Therefore, in practical phenomenological applications,  one often needs UPDFs for both gluons and quarks over a broad kinematic range. This has motivated the use of alternative UPDF models based on the DGLAP evolution equations, such as the Kimber-Martin-Ryskin (KMR)~\cite{KMR}, Martin-Ryskin-Watt (MRW)~\cite{MRW}, and parton branching (PB)~\cite{PB1,PB2} approaches. The $k_t$-factorization framework and different UPDF models have been applied to many high-energy processes and have provided useful phenomenological descriptions; see, for example, Refs.~\cite{PBDrell,Lipatov:2012td,KordValeshabadi:2021ybd,Valeshabadi:2025hhr,Lipatov:2016wgr,Rezaie:2025hzt,Valeshabadi:2021twn}.

	A particularly relevant phenomenological motivation for retaining transverse momentum dependence in DPS is that distributions probing transverse momentum imbalance and angular correlations are often central to experimental extractions of the DPS contribution. Depending on the analysis, these observables were used at the Tevatron in four-jet and $\gamma+3$-jet final states~\cite{CDF:1993sbj,CDF:1997yfa,D0:2009apj}, in early LHC measurements of $W+2$-jet production~\cite{ATLAS:2013aph,CMS:2013huw}, and in a more recent four-jet analysis~\cite{CMS:2021lxi}. The repeated use of such transverse momentum sensitive distributions in DPS extractions provides important phenomenological motivation for a transverse momentum dependent treatment of DPS.

	This motivation is illustrated in inclusive four-jet production, where angular-correlation observables such as $\Delta S$ probe transverse momentum imbalance and azimuthal decorrelations among the jets~\cite{CMS:2013slh,CMS:2021lxi}. It was shown in Ref.~\cite{Kutak:2016mik} that the $k_t$-factorization approach gives a better description of the $\Delta S$ distribution than the LO collinear framework. A related motivation is provided by double open-charm production, $pp\to c\bar c c\bar c X$, and correlated $D$-meson pair production, which were studied in Ref.~\cite{vanHameren:2014ava} using a factorized DPS ansatz in which each $c\bar c$ subprocess was calculated within the $k_t$-factorization approach with UPDFs. In that case, observables such as azimuthal-angle correlations and invariant-mass distributions of the produced charm-meson pairs are sensitive to the transverse momenta entering the two hard scatterings. The results indicate that DPS gives the dominant contribution and provides a reasonable description of several measured correlation observables, although some discrepancies remain, reflecting the limitations of a simple product ansatz for two independent scatterings. These studies therefore motivate a more complete UDPDF-based description, in which the transverse momenta of the two active partons entering a given DPDF are kept explicit and the role of longitudinal two-parton correlations can be investigated beyond the conventional factorized approximation.

	A practical construction of such UDPDFs can be based on last step evolution models developed for single UPDFs. In the MRW UPDF approach, starting from input collinear distributions, the transverse momentum dependence is generated from the last step of QCD evolution, while Sudakov form factors account for the probability of no resolvable emission between the transverse momentum scale and the hard scale. The simplicity of this approach motivates an analogous formulation of UDPDFs from input collinear DPDFs. MRW-type UDPDFs have already been discussed in Ref.~\cite{Golec-Biernat:2016vbt}, and the gluon-gluon distribution $\mathcal F^{gg}$ was investigated numerically in Ref.~\cite{Elias:2017flu}. However, in the conventional LO-MRW formalism~\cite{MRW}, the treatment of the region $k_t<\mu_0$, where $\mu_0$ denotes the initial scale of the evolution, requires special care. In the double parton case this leads to a piecewise structure, since different expressions may be needed depending on whether one or both transverse momenta lie below the input scale. The situation is particularly delicate for the non-homogeneous contribution, where the perturbative splitting scale introduces an additional scale and partially unintegrated terms are not straightforward to use in an off-shell $k_t$-factorized matrix element. These features make the direct LO-MRW (DLO-MRW) construction cumbersome for phenomenological applications of DPS in the $k_t$-factorization framework, especially when one aims to use the full flavor structure of non-factorized DPDFs.
		
	To avoid these difficulties, we consider MRW-type prescriptions that can be applied directly to the full collinear DPDF. In particular, we study the virtuality ordered MRW model (VO-MRW), motivated by the virtuality ordering used in the MRW framework~\cite{MRW}, and the modified KMRW model (MKMRW)~\cite{Guiot:2022psv}, which was introduced to address normalization ambiguities of conventional KMRW-type UPDFs. We also consider a normalization-matched version of the VO-MRW construction, denoted by MVO-MRW, in which the virtuality ordered transverse momentum shape is preserved while the integrated distribution is matched to the corresponding collinear DPDF. These prescriptions allow us to study the full flavor structure of non-factorized UDPDFs, including channels beyond the gluon-gluon case, while avoiding the piecewise low-$k_t$ treatment of the direct LO-MRW UDPDF model~\cite{Golec-Biernat:2016vbt,Elias:2017flu}. However, we also perform a numerical analysis of the LO-MRW UDPDF in the perturbative region $k_t\geq \mu_0$, where the last step expressions for the homogeneous and non-homogeneous contributions have the same structure and can be applied to the full collinear DPDF.

	The main purpose of this work is to perform a numerical investigation of UDPDFs constructed from non-factorized collinear DPDFs using the DLO-MRW, DMKMRW, DVO-MRW, and MDVO-MRW prescriptions. As input, we use the \texttt{GS09} DPDFs~\cite{Gaunt:2009re}, which are evolved to unequal scales using our numerical double DGLAP evolution program. The details of this program will be presented elsewhere. In the present paper, we summarize the ingredients needed for the construction of the UDPDFs and validate the unequal-scale DPDFs by testing the momentum and valence-number sum rules. We also examine the normalization properties of the different UDPDF prescriptions and study the impact of non-factorized DPDF correlations on the generated transverse momentum dependence.
	
	The UDPDFs considered in this work have the generic form $\mathcal{F}_{i,j}
	(x_1,x_2,k_{1t}^2,k_{2t}^2,\mu_A^2,\mu_B^2)$, where $k_{1t}$ and $k_{2t}$ denote the transverse momenta of the two active partons entering the two hard subprocesses. In order to quantify the effect of double DGLAP induced correlations, we compare non-factorized UDPDFs with the commonly used factorized approximation in different plots. This ratio allows us to study how longitudinal correlations generated by double DGLAP evolution are reflected in the unintegrated distributions. We also compare the different UDPDF prescriptions with each other in order to clarify the impact of the chosen last step evolution model on  phenomenological applications of DPS in the $k_t$-factorization framework.

	The paper is organized as follows. In Sec.~\ref{sec:ddglap}, we summarize the double DGLAP evolution equations and describe the unequal-scale numerical solution used in this work. In Sec.~\ref{sec:udpdfs}, we introduce the LO-MRW approach and then present the VO-MRW, MVO-MRW, and MKMRW models of UDPDFs, and briefly discuss the possible extension of the present construction to the $b$-dependent case. In Sec.~\ref{sec:results_discussion}, we validate the unequal-scale DPDFs using the momentum and valence-number sum rules, test the normalization properties of the UDPDFs, and study the transverse momentum dependence and non-factorized to factorized ratios for different channels and models. Finally, in Sec.~\ref{sec:conclusion}, we summarize our main findings.
	
	\section{Double DGLAP evolution equations}
	\label{sec:ddglap}
	
	The scale dependence of DPDFs is governed by evolution equations analogous to the ordinary DGLAP equations for SPDFs. These equations are usually referred to as double DGLAP evolution equations. In contrast to the SPDF case, the double DGLAP equations contain three distinct contributions. The first two terms describe the independent evolution of each of the two partons along its corresponding parton ladder. The third term is a non-homogeneous contribution, often called the feed term, and describes the perturbative splitting of one parent parton into the two observed daughter partons. At leading-order, the equal-scale double DGLAP evolution equation can be written as~\cite{Gaunt:2009re,Ceccopieri:2010kg}
	\begin{align}
		\label{eq_dDGLAP}
		\mu^2 \frac{dD_{ij}(x_1,x_2,\mu^2)}{d\mu^2}
		=
		\frac{\alpha_s(\mu^2)}{2\pi}
		\Biggl[
		&\sum_{k}\int_{x_1}^{1-x_2}\frac{dy_1}{y_1}
		D_{kj}(y_1,x_2,\mu^2)
		P_{k\to i}\left(\frac{x_1}{y_1}\right)
		\nonumber\\
		+
		&\sum_{l}\int_{x_2}^{1-x_1}\frac{dy_2}{y_2}
		D_{il}(x_1,y_2,\mu^2)
		P_{l\to j}\left(\frac{x_2}{y_2}\right)
		\nonumber\\
		+
		&\sum_k
		D_k(x_1+x_2,\mu^2)
		\frac{1}{x_1+x_2}
		P^R_{k\to ij}
		\left(\frac{x_1}{x_1+x_2}\right)
		\Biggr].
	\end{align}
	
	Here $D_{ij}(x_1,x_2,\mu^2)$ denotes the collinear DPDF for finding two partons of flavors $i$ and $j$, carrying longitudinal momentum fractions $x_1$ and $x_2$, at the common scale $\mu^2$. The functions $P_{k\to i}$ and $P_{l\to j}$ are the usual DGLAP splitting functions, including virtual corrections. The last term contains the real splitting function $P^R_{k\to ij}$, which describes the perturbative splitting of a single parent parton $k$ into the two daughter partons $i$ and $j$. Since this term corresponds to a real branching contribution, it does not contain the virtual parts of the splitting functions; see Ref.~\cite{Gaunt:2009re} for a detailed discussion.
	
	A numerical solution of the leading-order equal-scale double DGLAP equations was presented in Ref.~\cite{Gaunt:2009re}. In that work, authors proposed momentum and valence quark number sum rules for DPDFs and showed that, if these sum rules are imposed at the input scale, they are preserved by leading-order double DGLAP evolution. A general all-order proof of the sum rules, including the treatment of bare and renormalized distributions, was later given in Ref.~\cite{Diehl:2018kgr}. 
	
	The sum rules proposed in Ref.~\cite{Gaunt:2009re} express the fact that, once one parton has been observed, the momentum and valence content available to the remaining partons are constrained. The momentum sum rule reads
	\begin{equation}
		\label{eq_dpdf_momentum_sum_rule}
		\sum_{j_1}
		\int_0^{1-x_2} dx_1\,
		x_1\,D_{j_1j_2}(x_1,x_2,\mu^2)
		=
		(1-x_2)\,D_{j_2}(x_2,\mu^2).
	\end{equation}
	This relation states that, after selecting a parton of flavor $j_2$ with momentum fraction $x_2$, the remaining partons can carry only the momentum fraction $1-x_2$. The corresponding valence-number sum rule is
	\begin{equation}
		\label{dpdf_number_sum_rule}
		\int_0^{1-x_2} dx_1\,
		D_{j_{1v}j_2}(x_1,x_2,\mu^2)
		=
		\begin{cases}
			N_{j_{1v}}\,D_{j_2}(x_2,\mu^2),
			& j_2\neq j_1,\bar{j}_1, \\[0.2cm]
			\left(N_{j_{1v}}-1\right)D_{j_2}(x_2,\mu^2),
			& j_2=j_1, \\[0.2cm]
			\left(N_{j_{1v}}+1\right)D_{j_2}(x_2,\mu^2),
			& j_2=\bar{j}_1 .
		\end{cases}
	\end{equation}
	Here $j_{1v}\equiv j_1-\bar{j}_1$ denotes a valence combination, with $j_1\neq g$, and $N_{j_{1v}}$ is the corresponding valence quark number in the proton; for example, $N_{u_v}=2$ and $N_{d_v}=1$. The three cases in Eq.~\eqref{dpdf_number_sum_rule} reflect the change in the remaining net valence content after the second parton has been specified. If $j_2$ is neither $j_1$ nor $\bar{j}_1$, the full valence-number $N_{j_{1v}}$ remains available. If $j_2=j_1$, one quark of that flavor has already been selected, giving the factor $N_{j_{1v}}-1$. If $j_2=\bar{j}_1$, the net valence combination $j_1-\bar{j}_1$ is increased by one, giving the factor $N_{j_{1v}}+1$. The analogous relations with the two parton labels interchanged are obtained by integrating over $x_2$ instead of $x_1$. These sum rules therefore impose non-trivial constraints on the allowed initial DPDFs.

	Using these constraints, Ref.~\cite{Gaunt:2009re} constructed improved input DPDFs at $Q_0^2=1~\mathrm{GeV}^2$, based on the MSTW2008LO single PDF set~\cite{MSTW}. An important feature of the Gaunt-Stirling construction is that the input DPDFs are not taken to be simple products of two SPDFs. The starting distributions are first written in a pseudo-factorized form,
	\begin{equation}
		D_{ij}(x_1,x_2,Q_0^2)
		\simeq
		f_i(x_1,Q_0^2)\,
		f_j(x_2,Q_0^2)\,
		\rho_{ij}(x_1,x_2),
	\end{equation}
	where the phase-space factor $\rho_{ij}$ suppresses the distributions near the kinematic boundary $x_1+x_2=1$. This form is then modified by additional terms required by the number sum rules. Equal flavor valence-valence distributions receive a subtraction associated with the finite number of valence quarks in the proton, while quark-antiquark channels receive a positive correlation contribution related to the input sea distribution. These input correlations are important for the interpretation of the UDPDF ratios discussed in this work. In particular, the stronger suppression observed in the $uu$ channel is connected with valence-number constraints, whereas the enhancement that can appear in the $u\bar u$ channel reflects the quark-antiquark correlation term present already at the input scale. The equal-scale double DGLAP equations were then solved directly in $x$-space on a grid in the variable
	\begin{equation}
		u=\ln\left(\frac{x}{1-x}\right),
	\end{equation}
	which provides logarithmic resolution at small $x$ and also gives an appropriate treatment near the kinematic boundary $x_1+x_2=1$. The scale evolution was performed using a fourth-order Runge-Kutta algorithm. The convolution integrals involving regular, plus distribution, and delta function parts of the splitting functions were evaluated using Newton-Cotes quadrature with precomputed integration weights, while the SPDF feed terms were obtained by interpolating the simultaneously evolved single PDFs. The calculation was performed in a flavor basis formed by singlet, gluon, valence, and non-singlet tensor combinations, $\{\Sigma,g,V_i,T_i\}$, which reduces the number of coupled equations and simplifies the numerical evolution. The resulting distributions were made available through the \texttt{GS09} grid files and the corresponding Fortran interpolation routine~\cite{Gaunt:2009re}.

	For phenomenological applications, and in particular for the construction of UDPDFs, it is necessary to consider the more general case in which the two partons are probed at different scales. These unequal-scale DPDFs are denoted by $D_{ij}(x_1,x_2,\mu_1^2,\mu_2^2)$.
	Following Refs.~\cite{Gaunt:2009re,Ceccopieri:2010kg}, unequal-scale DPDFs can be obtained from the equal-scale distributions by first evolving both partons up to the lower of the two scales and then evolving only the parton associated with the higher scale. For example, for $\mu_2^2>\mu_1^2$, one first determines $D_{ij}(x_1,x_2,\mu_1^2,\mu_1^2)$ from the equal-scale double DGLAP equation and then uses it as the initial condition for the single-leg evolution equation
	\begin{align}
		\label{eq_unequalScale_dDDGLAP}
		\mu_2^2
		\frac{dD_{ij}(x_1,x_2,\mu_1^2,\mu_2^2)}
		{d\mu_2^2}
		=
		\frac{\alpha_s(\mu_2^2)}{2\pi}
		\sum_{l}
		\int_{x_2}^{1-x_1}
		\frac{dy_2}{y_2}
		D_{il}(x_1,y_2,\mu_1^2,\mu_2^2)
		P_{l\to j}
		\left(\frac{x_2}{y_2}\right).
	\end{align}
	For the opposite ordering, $\mu_1^2>\mu_2^2$, the corresponding equation is obtained by interchanging the two parton labels and evolving the first parton from $\mu_2^2$ to $\mu_1^2$. The non-homogeneous splitting contribution does not appear in Eq.~\eqref{eq_unequalScale_dDDGLAP}, because the perturbative splitting that produces the two daughter partons is already included in the equal-scale distribution at the lower scale. The subsequent evolution from the lower scale to the higher scale acts only on the parton associated with the higher hard subprocess.
		
	In this work, we solve Eq.~\eqref{eq_unequalScale_dDDGLAP} numerically using a method similar in spirit to that of Ref.~\cite{Gaunt:2009re}. The convolution integrals are evaluated in $x$-space, while the evolution in $t=\ln\mu^2$ is performed using the adaptive Runge-Kutta-Fehlberg (RKF45) algorithm implemented in the \texttt{GSL} library~\cite{GSLRefManual3rd}. The absolute and relative integration tolerances are set to $\epsilon_{\mathrm{abs}}=10^{-10}$ and $\epsilon_{\mathrm{rel}}=10^{-8}$, respectively. The longitudinal grid covers $10^{-6}\leq x<1$, while the scale range is $1~\mathrm{GeV}^2\leq\mu^2\leq10^9~\mathrm{GeV}^2$, in accordance with the GS09 DPDF set.
	
	The production calculation used for all subsequent UDPDF results employs $N_x^{\mathrm{solver}}=90$ longitudinal points in the internal evolution grid. The resulting DPDF tables are stored on a reduced grid containing $N_x^{\mathrm{stored}}=56$ longitudinal points and $N_t=41$ points. For the sum rule convergence study presented in Sec.~\ref{sec:results_discussion}, we use a refined configuration with $N_x^{\mathrm{solver}}=150$, $N_x^{\mathrm{stored}}=70$, and $N_t=60$. Here, $N_t$ denotes the number of stored scale-grid points, whereas the internal RKF45 integration steps are selected adaptively.
	
	 To improve numerical performance, the calculation is parallelized using \texttt{OpenMP}~\cite{dagum1998openmp}. We refer to our numerical implementation as \texttt{ChromaPDFEvolver}. A detailed description of the code and its architecture will be presented elsewhere. In the present work, we use it to generate unequal-scale DPDF grids needed for the construction of the unintegrated distributions. These grids are then read and interpolated with \texttt{PDFxTMDLib}~\cite{Valeshabadi2026}, which allows the evolved DPDFs to be used in phenomenological applications and in the UDPDF approaches discussed in the following section.

	\section{Unintegrated double parton distribution functions}
	\label{sec:udpdfs}
	
	In this section we define the UDPDFs used in the present work. The collinear DPDF is written as
	\begin{equation}
		\label{eq:DPDF-h-nh-short}
		D_{ij}(x_1,x_2,\mu_1^2,\mu_2^2)
		=
		D^{(h)}_{ij}(x_1,x_2,\mu_1^2,\mu_2^2)
		+
		D^{(nh)}_{ij}(x_1,x_2,\mu_1^2,\mu_2^2),
	\end{equation}
	where the homogeneous term corresponds to independent evolution of the two parton
	ladders, while the non-homogeneous term is generated by perturbative splitting of one
	parent parton into two daughter partons. In the numerical analysis of this work we use the full evolved DPDF, $D_{ij}=D^{(h)}_{ij}+D^{(nh)}_{ij}$,
	and do not separate the two components. This choice is important for the construction of
	the UDPDFs. The original KMR/MRW-type extension to double distributions requires a
	separate treatment of the homogeneous and non-homogeneous contributions in $k_t < \mu_0$. However, other approaches like virtuality-based, and MKMRW-inspired models do not have this limitation. 
	
	\subsection{MRW UPDFs and the limitations of the direct LO-MRW double distribution}
	\label{subsec:MRW_limitation}
	The MRW prescription constructs a UPDF from an
	ordinary collinear PDF by evolving collinear PDFs to the last step of the DGLAP evolution ladder. In the last step evolution ladder, the evolving parton emits a real parton, and then evolve to the scale $\mu$ with no real emission using the Sudakov form factor. The LO-MRW UPDF may be written as
	\begin{equation}
		\label{eq:LOMRW-UPDF-short}
		\mathcal F_i^{\mathrm{LO-MRW}}(x,k_t^2,\mu^2)
		=
		T_i(k_t^2,\mu^2)
		\frac{\alpha_s(k_t^2)}{2\pi k_t^2}
		\sum_j
		\int_x^1 \frac{dz}{z}\,
		P^{R}_{j \to i}(z)\,
		f_j\left(\frac{x}{z},k_t^2\right)
		\Theta_{ij}^{\mathrm{Ordering}}(z,k_t,\mu).
	\end{equation}
	In the above equation $P^{R}_{j \to i}$ is the LO splitting function, and $T_i$ is the Sudakov form factor,
	\begin{equation}
		\label{eq:MRW_Sudakov}
		T_i(k_t^2,\mu^2)=
		\begin{cases}
			\exp\left[
			-\int_{k_t^2}^{\mu^2}
			\frac{d\kappa_t^2}{\kappa_t^2}
			\frac{\alpha_s(\kappa_t^2)}{2\pi}
			\sum_{j}
			\int_0^1
			d\xi\,
			\xi\,P^{R}_{i \to j}(\xi)\,
			\Theta_{ij}^{\mathrm{Ordering}}(\xi,\kappa_t,\mu)
			\right], & k_t \leq \mu,\\[1mm]
			1, &  k_t > \mu,
		\end{cases}
	\end{equation}
	and $\Theta_{ij}^{\mathrm{Ordering}}$ implements the real emission
	cutoff. 
	In the MRW prescription, the cutoff is applied only to the diagonal splitting kernels, which contain soft gluon singularities,
	\begin{equation}
		\label{eq:MRWtheta_short}
		\Theta_{ij}^{\mathrm{Ordering}}(z,k_t,\mu)
		=
		\begin{cases}
			\Theta\left(z_{\max}(k_t,\mu)-z\right), & i=j,\\[1mm]
			1, & i\neq j .
		\end{cases}
	\end{equation}
The explicit form of $z_{\max}$ depends on the ordering condition imposed at the last step of the evolution. For angular ordering, one imposes $\tilde q_{i+1}>z_i\tilde q_i$
at the (i)th step of the evolution ladder~\cite{Valeshabadi:2021smo}. Applying this condition to the final branching gives~\cite{Valeshabadi:2021smo} $\mu > z \tilde q$, where $\tilde q = \frac{k_t}{1-z}$ is the rescaled transverse momentum.
Solving the inequality gives the angular ordering upper cutoff
\begin{equation}
	\label{eq:AOCutoff}
	z<z_{\max}^{\mathrm{AO}}(k_t,\mu) =\frac{\mu}{\mu+k_t}.
\end{equation}

For strong ordering constraint, the ordering is instead imposed directly on the rescaled transverse momentum, $\tilde q_{i+1}>\tilde q_i$ ~\cite{Valeshabadi:2021smo}. At the final branching this condition gives $\mu>\tilde q$. Therefore, one obtains~\cite{KimberThesis,Kimber:1999xc}
\begin{equation}
	\label{eq:SOCutoff}
	z<z_{\max}^{\mathrm{SO}}(k_t,\mu) =1-\frac{k_t}{\mu}.
\end{equation}
For $k_t\geq\mu$, the strong ordering cutoff satisfies $z_{\max}^{\mathrm{SO}}\leq 0$. Since the physical real emission region requires $0<z<1$, no allowed $z$ interval remains for the soft gluon emission, to which the strong ordering cutoff is applied. These contributions therefore vanish for transverse momenta at or above the hard scale. By contrast, $z_{\max}^{\mathrm{AO}}$ remains positive, so the angular ordering prescription can generate such contributions above $\mu$.  Moreover, this cutoff agrees with the $k_t \ll \mu$ expansion of the angular ordering cutoff,
$$
z_{\max}^{\mathrm{AO}}=\frac{\mu}{\mu+k_t}
=1-\frac{k_t}{\mu}+\mathcal{O}\left(\frac{k_t^2}{\mu^2}\right),
$$
up to first order in $k_t/\mu$.

Having specified the ordering constraint entering the real emission term, it is useful to note that, as shown in Ref.~\cite{MRW}, Eq.~\eqref{eq:LOMRW-UPDF-short} gives the integral form of the LO-MRW prescription and can be derived from the differential form
	\begin{equation}
		\label{eq:MRWDiff}
		\mathcal F_i^{\mathrm{LO-MRW}}(x,k_t^2,\mu^2)
		=
		\frac{\partial}{\partial k_t^2}
		\left[
		T_i(k_t^2,\mu^2)\,
		f_i(x,k_t^2)
		\right].
	\end{equation}
	This differential form is one of the motivations for the MKMRW method discussed below. However, it has been shown that the differential and integral forms do not always lead to identical distributions, and methods to fix this ambiguity is proposed in the Refs.~\cite{Golec-Biernat:2018hqo,Valeshabadi:2021smo,Guiot:2019vsm}.
	
	The LO-MRW expressions in Eqs.~\eqref{eq:LOMRW-UPDF-short} and \eqref{eq:MRWDiff} require special treatment for $k_t < \mu_0$, where $\mu_0$ is the initial factorization scale of the input collinear PDF. Since the collinear PDFs are not defined below $\mu_0$, Ref.~\cite{MRW} imposed the normalization condition
	\begin{equation}
		\label{eq:LOMRW_norm}
		\int_0^{\mu^2} dk_t^2\,
		\mathcal F_i(x,k_t^2,\mu^2)
		=
		f_i(x,\mu^2),
	\end{equation}
	and introduced the low-$k_t$ contribution which satisfies the normalization condition, i.e.
	\begin{equation}
		\label{eq:LOMRW_smallkt}
		\mathcal F_i^{\mathrm{LO-MRW}}(x, k_t^2 < \mu_0^2, \mu^2)
		=
		\frac{1}{\mu_0^2}
		f_i(x, \mu_0^2)\,
		T_i(\mu_0^2,\mu^2).
	\end{equation}
	The extension of this last step approach to DPDF is more
	restrictive than in the SPDF case. For the homogeneous part, the direct LO-MRW UDPDFs unfolds the last step independently on the two parton ladders. In the region in which both transverse momenta are perturbative, its schematic structure is
	\begin{equation}
		\label{eq:LOMRW-UDPDF-h-schematic}
		\mathcal F^{(h)}_{ij}
		\sim
		T_i(k_{1t}^2,\mu_1^2)\,
		T_j(k_{2t}^2,\mu_2^2)\,
		\frac{\alpha_s(k_{1t}^2)}{2\pi k_{1t}^2}
		\frac{\alpha_s(k_{2t}^2)}{2\pi k_{2t}^2}
		\,
		P\otimes P\otimes
		D^{(h)}(k_{1t}^2,k_{2t}^2).
	\end{equation}
	As seen in Eq. \eqref{eq:LOMRW_smallkt}, additional formulas are needed when one or both transverse momenta are below the input scale.
	
	The non-homogeneous contribution is even more delicate. In this case the perturbative
	splitting scale $\mu_s$ introduces an additional scale. Only the term in which both daughter partons subsequently undergo a resolvable last step
	branching, so that both $k_{1t}$ and $k_{2t}$ remain explicit, is suitable for use with
	off-shell $k_t$-factorized matrix elements. Schematically,
	\begin{equation}
		\label{eq:LOMRW-UDPDF-nh-acceptable}
		\mathcal F^{(nh)}_{ij}
		\sim
		T_i(k_{1t}^2,\mu_1^2)\,
		T_j(k_{2t}^2,\mu_2^2)\,
		\frac{\alpha_s(k_{1t}^2)}{2\pi k_{1t}^2}
		\frac{\alpha_s(k_{2t}^2)}{2\pi k_{2t}^2}
		\,
		P\otimes P\otimes
		D^{(nh)}(k_{1t}^2,k_{2t}^2).
	\end{equation}
	Terms in which one or both daughter partons do not undergo the final resolvable branching are not used here. Conceptually, in such terms there is an additional scale where splitting occurs $\mu_s \geq \mu_0$, and hence the two generated final state partons from single SPDF feed term must have $k_t \geq \mu_0$. Therefore, if one of the two final state partons has $k_t <\mu_0$ the UDPDF is not physically logical. This is the reason why the partially unintegrated non-homogeneous terms as discussed in Ref.~\cite{Golec-Biernat:2016vbt}, are not appropriate for the $k_t$-factorized calculation.
	
	Thus, the direct LO-MRW double construction is not a single compact expression for the full DPDF. It is a piecewise prescription that treats the homogeneous and non-homogeneous contributions separately. Moreover, if the single PDF entering the feed term is also made transverse momentum dependent, one must introduce transverse momentum dependent splitting functions, which considerably complicates the formalism~\cite{Golec-Biernat:2016vbt}. This part was also not included in the numerical study of Ref.~\cite{Elias:2017flu}. We therefore do not include such terms in the present VO-MRW, MVO-MRW, and MKMRW UDPDF models.
		
	For the present analysis, the need to separate $D^{(h)}$ and $D^{(nh)}$ for $k_t<\mu_0$ in the direct LO-MRW approach represents a practical limitation. Our collinear input is the full distribution, $D=D^{(h)}+D^{(nh)}$, while the two components are not evolved and stored separately. Therefore, in the numerical analysis, we use the direct LO-MRW UDPDF only for $k_t\geq\mu_0$. By contrast, the virtuality ordered MRW model introduced below, together with its normalization matched version and the normalized MKMRW prescription, can be applied directly to the full collinear DPDF. These prescriptions therefore avoid the need for partially unintegrated non-homogeneous contributions over the entire $k_t$ range.
	
	\subsection{Virtuality ordered MRW construction for UPDFs and UDPDFs}
	\label{subsec:VOMRW_UDPDFs}
	
	We first consider a virtuality ordered variant of the MRW prescription, denoted by VO-MRW. This prescription is motivated by the NLO formulation of the MRW approach~\cite{MRW}, in which the scale entering the collinear distribution, the coupling, and the Sudakov form factor is the virtuality of the parton in the last branching rather than the transverse momentum itself. For a last step splitting with longitudinal momentum fraction \(z\), this virtuality is taken to be
	\begin{equation}
		\label{eq:VOMRW_virtuality}
		K^2(z,k_t^2)
		=
		\frac{k_t^2}{1-z}.
	\end{equation}
	In the present work we use this virtuality ordered kinematics together with LO splitting functions. Thus the label VO-MRW refers to the ordering variable used in the last step.
	
	With the same convention as above, the VO-MRW UPDF is written as
	\begin{equation}
		\label{eq:VO-MRW-UPDF}
		\begin{aligned}
			\mathcal F_i^{\mathrm{VO-MRW}}(x,k_t^2,\mu^2)
			&=
			\sum_{j}
			\int_x^1 \frac{dz}{z}\,
			T_i\left(K^2(z,k_t^2),\mu^2\right)
			\frac{\alpha_s\left(K^2(z,k_t^2)\right)}{2\pi k_t^2}
			\\
			&\quad \times
			P_{j \to i}^{R}(z)\,
			f_j\left(\frac{x}{z},K^2(z,k_t^2)\right)\,
			\Theta_{ij}^{\mathrm{Ordering}}(z,k_t,\mu)
			\\
			&\quad \times
			\Theta\left(\mu^2-K^2(z,k_t^2)\right)
			\Theta\left(K^2(z,k_t^2)-\mu_0^2\right).
		\end{aligned}
	\end{equation}
	The two theta functions $\Theta\left(\mu^2-K^2(z,k_t^2)\right)$ and $\Theta\left(K^2(z,k_t^2)-\mu_0^2\right)$, impose virtuality ordering and the requirement that the
	collinear PDF is evaluated only in the perturbative region, respectively, i.e.:
	\begin{equation}
		K^2(z,k_t^2)\leq \mu^2,
		\qquad
		K^2(z,k_t^2)\geq \mu_0^2 .
	\end{equation}
	Since the scale $K^2$ depends on $z$, both the Sudakov form factor and the running coupling must remain inside the $z$-integral.

	The VO-MRW UPDF has two useful consequences. First, because
	$K^2=k_t^2/(1-z)\geq k_t^2$, the condition $K^2<\mu^2$ implies
	$k_t^2<\mu^2$. Therefore, unlike the angular ordering version of conventional LO-MRW, the VO-MRW distribution does not generate a tail above the hard scale $\mu$.
	Second, the distribution can still be nonzero for $k_t^2<\mu_0^2$, provided that the integration over $z$ contains values for which
	$K^2(z,k_t^2)>\mu_0^2$. Thus no separate constant extrapolation similar to the Eq. \eqref{eq:LOMRW_smallkt} below
	$\mu_0$ is required, and the low $k_t$ contribution is generated dynamically, subject to the constraints in Eq.~\eqref{eq:VO-MRW-UPDF}.
	
	The same virtuality ordered last step prescription can be applied to double parton distributions. For the two final branchings we define
	\begin{equation}
		\label{eq:VOMRW_double_virtualities}
		K_1^2(z_1,k_{1t}^2)
		=
		\frac{k_{1t}^2}{1-z_1},
		\qquad
		K_2^2(z_2,k_{2t}^2)
		=
		\frac{k_{2t}^2}{1-z_2}.
	\end{equation}
	The corresponding VO-MRW UDPDF is
	\begin{equation}
		\label{eq:VO-MRW-UDPDF}
		\begin{aligned}
			\mathcal F_{ij}^{\mathrm{DVO-MRW}}
			& (x_1,x_2,k_{1t}^2,k_{2t}^2,\mu_1^2,\mu_2^2)
			\\
			&=
			\sum_{k,l}
			\int_{\frac{x_1}{1-x_2}}^{1}
			\frac{dz_1}{z_1}
			\int_{\frac{x_2}{1-x_1/z_1}}^{1}
			\frac{dz_2}{z_2}\,
			T_i\left(K_1^2,\mu_1^2\right)\,
			T_j\left(K_2^2,\mu_2^2\right)
			\\
			&\quad \times
			\frac{\alpha_s(K_1^2)}{2\pi k_{1t}^2}
			\frac{\alpha_s(K_2^2)}{2\pi k_{2t}^2}\,
			P_{k \to i}^{R}(z_1)\,
			P_{l \to j}^{R}(z_2)
			\\
			&\quad \times
			D_{kl}\left(
			\frac{x_1}{z_1},
			\frac{x_2}{z_2},
			K_1^2,
			K_2^2
			\right)
			\Theta_{ik}^{\mathrm{Ordering}}(z_1,k_{1t},\mu_1)\,
			\Theta_{jl}^{\mathrm{Ordering}}(z_2,k_{2t},\mu_2)
			\\
			&\quad \times
			\Theta(\mu_1^2-K_1^2)\,
			\Theta(\mu_2^2-K_2^2)\,
			\Theta(K_1^2-\mu_0^2)\,
			\Theta(K_2^2-\mu_0^2),
		\end{aligned}
	\end{equation}
	where $K_1^2\equiv K_1^2(z_1,k_{1t}^2)$ and
	$K_2^2\equiv K_2^2(z_2,k_{2t}^2)$. The lower limits in the $z_1$ and
	\(z_2\) integrals enforce the DPDF support condition
	\begin{equation}
		\frac{x_1}{z_1}+\frac{x_2}{z_2}\leq 1 .
	\end{equation}
	
	In Eq.~\eqref{eq:VO-MRW-UDPDF}, $D_{kl}$ denotes the full collinear DPDF of Eq.~\eqref{eq:DPDF-h-nh-short}.
	Therefore the VO-MRW model can be applied directly to the DPDF used in the present work, without separating the homogeneous and non-homogeneous components. Also, similar to what is already discussed, We do not include terms in which the feed term or non-homogeneous term become $k_t$ dependent in the MRW sense. 
	
	The VO-MRW method therefore provides a single compact expression for the UDPDFs over the relevant transverse momentum range. In contrast to the direct LO-MRW UDPDF, no separate formula is needed for $k_{t}<\mu_0$, and no explicit separation of
	$D^{(h)}$ and $D^{(nh)}$ is required. This makes the formulation simpler and better suited to the full DPDF study used in this work. The price is that the scale entering the DPDF, the coupling and the Sudakov factors is $z$-dependent, so these quantities cannot be taken outside the longitudinal momentum integrals, which leads to significant computational overhead. It should be noted that the normalization constraint in the VO-MRW model should satisfy Eq.~\eqref{eq:LOMRW_norm} similar to the LO-MRW/KMR UPDF model. However, the VO-MRW prescription is not constructed as an exact normalization-preserving ansatz, rather, it is a virtuality ordered last step model for generating the transverse momentum dependence from the full collinear DPDF.
	
	\subsection{Normalization Matched virtuality ordered MRW construction for UPDFs and UDPDFs}
	\label{subsec:matched_VOMRW_UDPDFs}
	The virtuality ordered MRW construction introduced in the previous subsection provides a compact last step model for generating the transverse momentum dependence of UPDFs and UDPDFs. However, unlike the MKMRW model, this model is not constructed to satisfy an exact normalization condition after integration over transverse momentum. In particular, because of the virtuality ordering constraint, the angular ordering cutoff, and the finite phase space available for the last branching, the integral of the VO-MRW unintegrated distribution over $k_t^2$ does not necessarily reproduce the corresponding collinear distribution exactly. This can introduce an additional normalization ambiguity when comparing different UPDF and UDPDF models or when using them in phenomenological applications.
	
	To reduce this ambiguity, we introduce a normalization matched version of the VO-MRW (MVO-MRW) approach. The idea is to preserve the transverse momentum shape generated by the virtuality ordered last step, while imposing the correct collinear normalization by a multiplicative matching factor. 
	
	For fixed-$x$ and $\mu^2$, the allowed transverse momentum range is limited by the virtuality ordering condition. Since $z\geq x$ and
	$$
	K^2=\frac{k_t^2}{1-z}<\mu^2 ,
	$$
	the maximal transverse momentum satisfies
	\begin{equation}
		\label{eq:matched_vomrw_single_ktmax}
		k_t^2
		<
		k_{t,\max}^2(x,\mu^2)
		=
		\mu^2(1-x).
	\end{equation}
	In the numerical implementation the lower limit is taken to be a small positive cutoff $k_{t,\min}^2$, introduced only to avoid the endpoint $k_t^2=0$, where in this work for our numerical calculation we set it $k_{t,\min}^2=10^{-6}$. The normalization integral of the unmatched shape is then
	\begin{equation}
		\label{eq:matched_vomrw_single_denominator}
		\mathcal N_i^{\mathrm{VO-MRW}}(x,\mu^2)
		=
		\int_{k_{t,\min}^2}^{k_{t,\max}^2(x,\mu^2)}
		d k_t^2\,
		\mathcal F_i^{\mathrm{VO-MRW}}(x,k_t^2,\mu^2).
	\end{equation}
	The matched single parton distribution is defined by
	\begin{equation}
		\label{eq:matched_vomrw_single}
		\mathcal F_i^{\mathrm{MVO-MRW}}(x,k_t^2,\mu^2)
		=
		\mathcal C_i^{\mathrm{VO-MRW}}(x,\mu^2)\,
		\mathcal F_i^{\mathrm{VO-MRW}}(x,k_t^2,\mu^2),
	\end{equation}
	where the matching coefficient is
	\begin{equation}
		\label{eq:matched_vomrw_single_factor}
		\mathcal C_i^{\mathrm{VO-MRW}}(x,\mu^2)
		=
		\frac{
			f_i(x,\mu^2)
		}{
			\mathcal N_i^{\mathrm{VO-MRW}}(x,\mu^2)
		}.
	\end{equation}
	With this definition one obtains
	\begin{equation}
		\label{eq:matched_vomrw_single_norm}
		\int_{k_{t,\min}^2}^{k_{t,\max}^2(x,\mu^2)}
		dk_t^2\,
		\mathcal F_i^{\mathrm{MVO-MRW}}(x,k_t^2,\mu^2)
		=
		f_i(x,\mu^2).
	\end{equation}
	Thus the MVO-MRW UPDF has the same transverse momentum shape as the original virtuality ordered construction, but its normalization in contrast to the simple VO-MRW UPDF model, is matched to the input collinear PDF.
	
	The same procedure can be applied to the double distribution.  The upper limits in the transverse momenta are fixed by the DPDF kinematic limit condition together with virtuality ordering. For the first leg,
	\[
	z_1\geq \frac{x_1}{1-x_2},
	\qquad
	K_1^2=\frac{k_{1t}^2}{1-z_1}<\mu_1^2,
	\]
	which gives
	\begin{equation}
		\label{eq:matched_dvomrw_kt1max}
		k_{1t,\max}^2
		=
		\mu_1^2
		\left(
		1-\frac{x_1}{1-x_2}
		\right).
	\end{equation}
	Similarly, for the second leg,
	\begin{equation}
		\label{eq:matched_dvomrw_kt2max}
		k_{2t,\max}^2
		=
		\mu_2^2
		\left(
		1-\frac{x_2}{1-x_1}
		\right).
	\end{equation}
	These limits vanish as the longitudinal configuration approaches the kinematic boundary $x_1+x_2=1$, as expected.
	
	The normalization denominator of the double shape is
	\begin{equation}
		\label{eq:matched_dvomrw_denominator}
		\begin{aligned}
			\mathcal N_{ij}^{\mathrm{DVO-MRW}}
			&(x_1,x_2,\mu_1^2,\mu_2^2)
			\\
			&=
			\int_{ k_{1t,\min}^2}^{k_{1t,\max}^2}
			dk_{1t}^2
			\int_{k_{2t,\min}^2}^{k_{2t,\max}^2}
			dk_{2t}^2\,
			\mathcal F_{ij}^{\mathrm{DVO-MRW}}
			(x_1,x_2,k_{1t}^2,k_{2t}^2,\mu_1^2,\mu_2^2).
		\end{aligned}
	\end{equation}
	The matched double distribution is then defined as
	\begin{equation}
		\label{eq:matched_dvomrw}
		\mathcal F_{ij}^{\mathrm{MDVO-MRW}}
		(x_1,x_2,k_{1t}^2,k_{2t}^2,\mu_1^2,\mu_2^2)
		=
		\mathcal C_{ij}^{\mathrm{DVO-MRW}}
		(x_1,x_2,\mu_1^2,\mu_2^2)\,
		\mathcal F_{ij}^{\mathrm{DVO-MRW}}
		(x_1,x_2,k_{1t}^2,k_{2t}^2,\mu_1^2,\mu_2^2),
	\end{equation}
	with
	\begin{equation}
		\label{eq:matched_dvomrw_factor}
		\mathcal C_{ij}^{\mathrm{DVO-MRW}}
		(x_1,x_2,\mu_1^2,\mu_2^2)
		=
		\frac{
			D_{ij}(x_1,x_2,\mu_1^2,\mu_2^2)
		}{
			\mathcal N_{ij}^{\mathrm{DVO-MRW}}(x_1,x_2,\mu_1^2,\mu_2^2)
		}.
	\end{equation}
	By construction, this gives
	\begin{equation}
		\label{eq:matched_dvomrw_norm}
		\begin{aligned}
			&\int_{k_{1t,\min}^2}^{k_{1t,\max}^2}
			dk_{1t}^2
			\int_{k_{2t,\min}^2}^{k_{2t,\max}^2}
			dk_{2t}^2\,
			\mathcal F_{ij}^{\mathrm{MDVO-MRW}}
			(x_1,x_2,k_{1t}^2,k_{2t}^2,\mu_1^2,\mu_2^2)
			\\
			&\qquad =
			D_{ij}(x_1,x_2,\mu_1^2,\mu_2^2).
		\end{aligned}
	\end{equation}
	Therefore, the normalization matched double VO-MRW (MDVO-MRW) prescription combines two useful features. It retains the virtuality ordered last step transverse momentum dependence of Eq.~\eqref{eq:VO-MRW-UDPDF}, but it also satisfies the collinear DPDF normalization condition. This makes it more suitable for phenomenological comparisons, because differences between models are less affected by an overall normalization mismatch and more directly reflect differences in the $k_t$-dependent shape. 
	
	\subsection{MKMRW approach for UPDFs and UDPDFs}
	\label{subsec:MKMRW_UDPDFs}
	The modified KMRW (MKMRW) prescription was proposed in
	Ref.~\cite{Guiot:2022psv} to address normalization ambiguities of the
	conventional KMRW formalism. In the standard LO-MRW approach, the UPDF is
	usually normalized by integrating the transverse momentum only up to the hard
	scale. In the angular ordering case, however, the UPDF can have support for
	transverse momenta larger than the hard scale, $k_t>\mu$, and the differential
	and integral definitions are not automatically equivalent~\cite{Golec-Biernat:2018hqo,Guiot:2019vsm,Valeshabadi:2021smo}.
	The MKMRW prescription avoids this ambiguity by imposing the normalization over
	the full transverse momentum range,
	\begin{equation}
		\label{eq:MKMRW_UPDF_norm}
		\int_0^\infty dk_t^2\,
		\mathcal F_i^{\mathrm{MKMRW}}(x,k_t^2,\mu^2)
		=
		f_i(x,\mu^2).
	\end{equation}
	
	In this method the collinear PDF is evaluated at the hard scale
	$\mu^2$. The transverse momentum dependence is generated entirely by a
	modified Sudakov form factor,
	\begin{equation}
		\label{eq:MKMRW_UPDF_diff}
		\mathcal F_i^{\mathrm{MKMRW}}(x,k_t^2,\mu^2)
		=
		\frac{\partial}{\partial k_t^2}
		\left[
		T_i^{\mathrm{MKMRW}}(k_t^2,\mu^2)\,
		f_i(x,\mu^2)
		\right].
	\end{equation}
	Since $f_i(x,\mu^2)$ is independent of $k_t^2$, the transverse distribution is
	controlled by the derivative of the Sudakov factor.
	
	The modified Sudakov factor is defined as
	\begin{equation}
		\label{eq:MKMRW_Sudakov}
		T_i^{\mathrm{MKMRW}}(k_t^2,\mu^2)
		=
		\exp\left[
		-\int_{k_t^2}^{\infty}
		\frac{d\kappa_t^2}{\kappa_t^2}
		\frac{\alpha_s(\kappa_t^2)}{2\pi}
		\sum_{k}
		\int_0^{z_{\max}(\kappa_t,\mu)}
		d\xi\,
		\xi\,P^{R}_{i \to k}(\xi)
		\right],
	\end{equation}
	where similar to the Ref.~\cite{Guiot:2022psv}, in the above formula to avoid Landau pole, the following saturated form of the QCD coupling constant is utilized:
	\begin{equation}
	\alpha_s(\mu^2) = \mathrm{Min}\bigg[\dfrac{12\pi}{(33-6)\ln(\dfrac{\mu^2}{\lambda_{QCD}^{2}})}, 0.4\bigg].
	\end{equation}
	An important difference from the ordinary MRW prescription is that, in the MKMRW approach, similar to the KMR approach~\cite{KMR}, the cutoff is imposed on all real splitting channels. Thus the same upper limit $z_{\max}$ is used for every channel in the sum over $k$, not only for the diagonal soft gluon kernels.
	
	For the angular ordering version used in this work, $z_{\max}$ is the same as in Eq.~\eqref{eq:AOCutoff}. With this definition the Sudakov factor satisfies
	\begin{equation}
		T_i^{\mathrm{MKMRW}}(\infty,\mu^2)=1,
		\qquad
		T_i^{\mathrm{MKMRW}}(0,\mu^2)=0,
	\end{equation}
	so that the normalization condition in Eq.~\eqref{eq:MKMRW_UPDF_norm} is
	satisfied.
	
	Taking the derivative in Eq.~\eqref{eq:MKMRW_UPDF_diff} gives
	\begin{equation}
		\label{eq:MKMRW_UPDF_integral}
		\mathcal F_i^{\mathrm{MKMRW}}(x,k_t^2,\mu^2)
		=
		f_i(x,\mu^2)\,
		\mathcal K_i^{\mathrm{MKMRW}}(k_t^2,\mu^2),
	\end{equation}
	where the normalized transverse kernel is
	\begin{equation}
		\label{eq:MKMRW_kernel}
		\mathcal K_i^{\mathrm{MKMRW}}(k_t^2,\mu^2)
		=
		\frac{\alpha_s(k_t^2)}{2\pi k_t^2}
		T_i^{\mathrm{MKMRW}}(k_t^2,\mu^2)
		\sum_{k}
		\int_0^{z_{\max}(k_t,\mu)}
		d\xi\,
		\xi\,P^R_{i \to k}(\xi).
	\end{equation}
	The same cutoff convention is used in the real emission kernel: all channels in the sum over $k$ are integrated only up to $z_{\max}(k_t,\mu)$. With the modified Sudakov factor, the transverse kernel is normalized over the full $k_t$ range,
	\begin{equation}
		\label{eq:MKMRW_kernel_norm}
		\int_0^\infty dk_t^2\,
		\mathcal K_i^{\mathrm{MKMRW}}(k_t^2,\mu^2)
		=
		1 .
	\end{equation}
	In the conventional LO-MRW/KMR construction, different expressions are required for $k_t<\mu_0$ and $k_t\geq \mu_0$. In the MKMRW prescription, as in the VO-MRW approach, this additional step is not needed. The transverse kernel $\mathcal K_i^{\mathrm{MKMRW}}$ is normalized over the full range $0<k_t<\infty$, so the low-$k_t$ contribution with $k_t<\mu_0$ is already included in the kernel normalization.
	
	We now use the normalized MKMRW kernel to construct a double unintegrated distribution. Unlike the direct LO-MRW UDPDF, and unlike the VO-MRW construction, this prescription does not unfold the last DGLAP step of the double distribution. Instead, it preserves the full collinear DPDF after integration over transverse momenta and assigns to each external parton a normalized MKMRW transverse kernel. We define
	\begin{equation}
		\label{eq:MKMRW_UDPDF}
		\begin{aligned}
			\mathcal F_{ij}^{\mathrm{DMKMRW}}
			& (x_1,x_2,k_{1t}^2,k_{2t}^2,\mu_1^2,\mu_2^2)
			\\
			&=
			D_{ij}(x_1,x_2,\mu_1^2,\mu_2^2)\,
			\mathcal K_i^{\mathrm{MKMRW}}(k_{1t}^2,\mu_1^2)\,
			\mathcal K_j^{\mathrm{MKMRW}}(k_{2t}^2,\mu_2^2).
		\end{aligned}
	\end{equation}
	Here $D_{ij}$ denotes the full collinear DPDF of Eq.~\eqref{eq:DPDF-h-nh-short}.
	Thus, as in the VO-MRW construction, the MKMRW model can be applied directly to the DPDF used in the present analysis, without separating the homogeneous and non-homogeneous components.
	
	By construction, the double distribution satisfies
	\begin{equation}
		\label{eq:MKMRW_UDPDF_norm}
		\int_0^\infty dk_{1t}^2
		\int_0^\infty dk_{2t}^2\,
		\mathcal F_{ij}^{\mathrm{DMKMRW}}
		(x_1,x_2,k_{1t}^2,k_{2t}^2,\mu_1^2,\mu_2^2)
		=
		D_{ij}(x_1,x_2,\mu_1^2,\mu_2^2).
	\end{equation}
	Similar to the VO-MRW model, the MKMRW construction avoids the complication associated with the low-$k_t$ treatment of the non-homogeneous UDPDF contribution in the direct LO-MRW approach. In particular, no separate piecewise prescription is required for the region $k_t<\mu_0$.
	
	This simplicity is useful for phenomenological applications. The same
	expression applies over the full transverse momentum range, the cutoff is applied independently to the two external parton legs through the kernels $\mathcal K_i^{\mathrm{MKMRW}}$ and $\mathcal K_j^{\mathrm{MKMRW}}$, and the normalization to the full collinear DPDF is exact. Consequently, the transverse dependence is factorized between the two parton legs and is not generated by an explicit last step convolution with the DPDF. Compared with the DVO-MRW and
	MDVO-MRW prescriptions, the MKMRW construction is also numerically less expensive and is therefore particularly convenient for phenomenological applications.
	
\subsection{Scope of the present UDPDF models and possible $b$-dependent extensions}
\label{subsec:b_dependent_extension}

	Although our numerical study uses the \texttt{GS09} DPDFs, the UDPDF prescriptions
	developed here are not tied to this specific input. In the present implementation,
	we use the \texttt{GS09} equal-scale DPDF grids as the collinear input, and the
	unequal-scale distributions required for the UDPDF calculation are generated
	through single-leg evolution. This choice fixes the longitudinal two-parton input
	and allows us to focus on the dependence of the results on the transverse momentum generation prescription.
	
	The formalism itself, however, can be interfaced with other DPDF inputs. Other
	equal-scale DPDF models could be incorporated by producing compatible numerical
	grids and using them in the same UDPDF prescriptions. For example, the recent $X$-ordered construction of Ref.~\cite{Fedkevych:2025lgp} modifies the PYTHIA-inspired multiparton PDF procedure in order to improve symmetry properties and the approximate fulfillment of the momentum and valence-number sum rules. Using such inputs would allow one to assess the uncertainty associated with the poorly constrained longitudinal DPDFs. In the present work, however, we keep the
	DPDF input fixed and non-factorized in order to isolate the dependence on the transverse momentum generation prescription.

	An important limitation of this work is its lack of consideration of the transverse interparton-distance dependence. The UDPDF models considered here are based on the factorized ansatz in Eq.~\eqref{eq:b_factorized_ansatz}, where the longitudinal DPDF and the transverse interparton-distance dependence are treated independently. As discussed in Refs.~\cite{Diehl:2017kgu,Diehl:2011yj}, this factorization can be violated by the non-homogeneous feed term associated with the perturbative splitting of one parent parton into the two observed partons. The DPDFs studied in the present work can therefore be considered as the integral of corresponding $b$-dependent DPDFs, with a suitably chosen cut-off. This interpretation applies in particular to the unequal-scale GS09-based distributions used in our numerical analysis. A more complete approach would therefore require suitable $b$-dependent DPDFs; for example, one may use the sum-rule-improved construction of Ref.~\cite{Diehl:2020xyg}. However, because such a formulation involves several transverse variables, such as $b$, $k_{1t}$, and $k_{2t}$, it is less straightforward than the $b$-integrated DPDF treatment adopted here.

	A related issue was discussed in Ref.~\cite{Elias:2017flu}, where the unintegrated double gluon distribution was studied with explicit dependence on the transverse momenta of the two active partons. In that work, however, the additional transverse momentum conjugate to the relative transverse separation of the two partons was set to zero, which corresponds to using distributions integrated over the relative transverse position. It was noted that this dependence could, in principle, be restored through a suitable form factor, although a realistic treatment of transverse two-parton correlations remains challenging~\cite{Elias:2017flu}. At a simplified model level, one possible extension would be to start from a $b$-dependent DPDF, such as the sum rule improved approach of Ref.~\cite{Diehl:2020xyg}, and apply the present transverse momentum generating prescriptions at fixed $b$. In this approach, the $b$-dependent DPDF would provide the dependence on $x_1$, $x_2$, and the interparton distance, while the MRW based kernels or last step evolution prescriptions would generate the dependence on $k_{1t}$ and $k_{2t}$. For example, the normalized kernel approach, i.e. DMKMRW, could be generalized schematically as
	\begin{equation}
		\mathcal F_{ij}
		(x_1,x_2,b,k_{1t}^2,k_{2t}^2,\mu_1^2,\mu_2^2)
		=
		D_{ij}(x_1,x_2,b,\mu_1^2,\mu_2^2)
		\mathcal K_i^{\mathrm{MKMRW}}(k_{1t}^2,\mu_1^2)
		\mathcal K_j^{\mathrm{MKMRW}}(k_{2t}^2,\mu_2^2).
	\end{equation}
	With normalized kernels, integration over $k_{1t}^2$ and $k_{2t}^2$ would reproduce the underlying $b$-dependent DPDF at fixed $b$.
	
	This extension would nevertheless remain a simplified model ansatz, since it assumes that the generated $k_t$ dependence factorizes from the $b$-dependence. It would therefore not capture possible genuine correlations among $x_1$, $x_2$, $b$, $k_{1t}$, and $k_{2t}$. A complete treatment of simultaneous $b$- and $k_t$-dependence is beyond the scope of the present work and is left for future study.

	\section{Results and Discussion}
	\label{sec:results_discussion}
	
	In this section we present the numerical results for the UDPDFs introduced in this work. We first validate the unequal-scale DPDFs used as input by testing the momentum and valence-number sum rules. We then examine the normalization properties of the UDPDF prescriptions and compare the transverse momentum dependence generated by the unintegrated DLO-MRW, DMKMRW, DVO-MRW, and MDVO-MRW models. The purpose of including MDVO-MRW is to separate effects due to the virtuality ordered transverse momentum shape from effects due to the overall normalization mismatch of the unmatched DVO-MRW prescription. The main goal is to determine in which kinematic regions the use of unintegrated non-factorized DPDFs is important and how strongly the result depends on the prescription used to generate the transverse momenta. Before presenting the results, we note the following setup. We adopt the MSTW2008LO PDF set~\cite{MSTW}, which is consistent with the \texttt{GS09} DPDF set, and access it through the PDFxTMDLib library~\cite{Valeshabadi2026}. In all UDPDF models considered below, we use the angular ordering cutoff in Eq.~\eqref{eq:AOCutoff}, following the standard choice in KMR/MRW-type prescriptions~\cite{KMR,MRW}. This cutoff implements the coherence constraint for soft gluon emissions and is used as the default ordering prescription in the present numerical analysis. The same analysis can also be performed with the strong ordering cutoff.

	Turning now to the validation step mentioned above, the momentum and valence-number sum rules for unequal-scale DPDFs provide useful consistency checks of the single-leg evolution. We first consider the momentum sum rule and then investigate the corresponding number sum rule.

	For the ordering $\mu_1^2>\mu_2^2$, where the first parton is evolved to the higher scale while the second parton is kept fixed at $\mu_2^2$, the unequal-scale momentum sum rule reads
	\begin{equation}
		\label{eq:unequal_scale_momentum_sum_rule}
		\sum_i
		\int_0^{1-x_2} dx_1\,x_1\,
		D_{ij}(x_1,x_2,\mu_1^2,\mu_2^2)
		=
		(1-x_2)\,f_j(x_2,\mu_2^2).
	\end{equation}
	This relation follows from the equal-scale Gaunt-Stirling momentum sum rule~\cite{Gaunt:2009re}. Once the second parton with momentum fraction $x_2$ has been selected, the remaining partons can carry only the momentum fraction $1-x_2$, which gives the prefactor on the right-hand side of Eq.~\eqref{eq:unequal_scale_momentum_sum_rule}.
	
	In the numerical validation we do not use the external SPDF directly as the reference value. Instead, we compare the unequal-scale momentum integral with the corresponding equal-scale DPDF momentum integral at the lower scale,
	\begin{equation}
		\label{eq:unequal_scale_momentum_numerical_test}
		R_j^M(x_2,\mu_1^2,\mu_2^2)
		=
		\frac{
			\displaystyle
			\sum_i
			\int_0^{1-x_2} dx_1\,x_1\,
			D_{ij}(x_1,x_2,\mu_1^2,\mu_2^2)
		}{
			\displaystyle
			\sum_i
			\int_0^{1-x_2} dx_1\,x_1\,
			D_{ij}(x_1,x_2,\mu_2^2,\mu_2^2)
		}.
	\end{equation}
	If the equal-scale DPDF satisfies the Gaunt-Stirling momentum sum rule exactly, the denominator of Eq.~\eqref{eq:unequal_scale_momentum_numerical_test} is equal to $(1-x_2)f_j(x_2,\mu_2^2)$. This ratio therefore tests whether the single-leg evolution from $\mu_2^2$ to $\mu_1^2$ preserves the momentum carried by the evolved parton sector, while reducing possible numerical mismatches between the DPDF interpolation and the external SPDF evaluation.
	
\begin{figure}[htbp]
	\centering
	\includegraphics[width=0.49\textwidth]{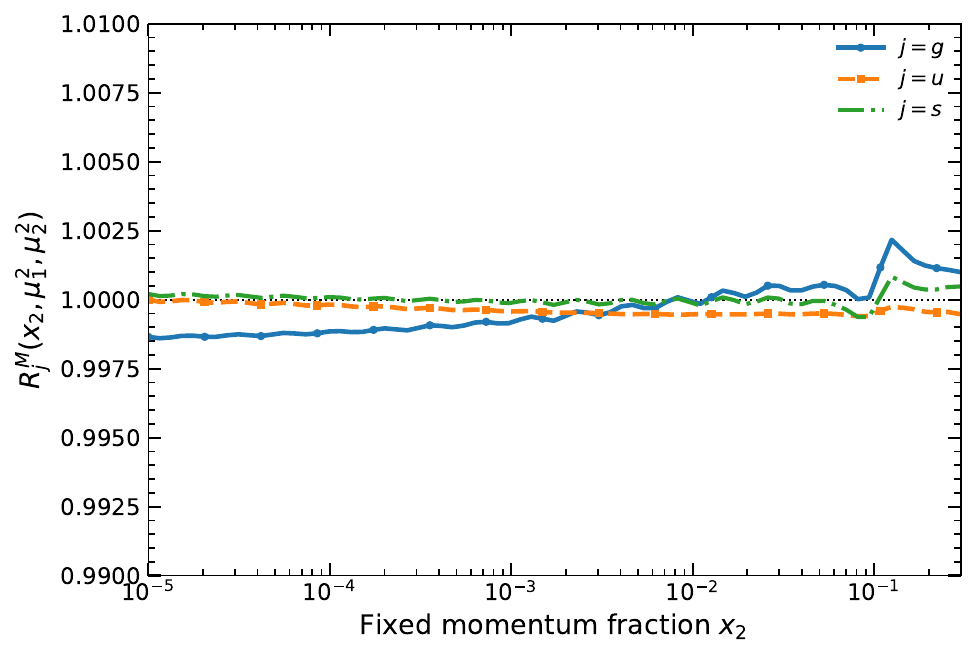}
	\includegraphics[width=0.49\textwidth]{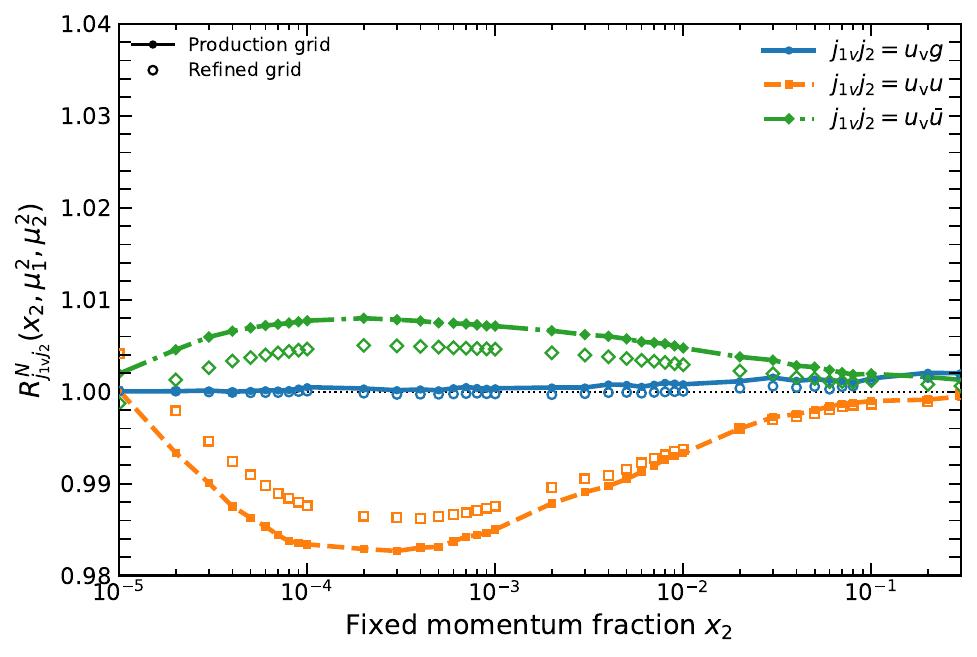}
\caption{
	Validation of the unequal-scale momentum and valence-number sum rules in \texttt{ChromaPDFEvolver}. Left panel: momentum sum rule ratio $R_j^M(x_2,\mu_1^2,\mu_2^2)$ defined in
	Eq.~\eqref{eq:unequal_scale_momentum_numerical_test}, shown for fixed second parton $j=g,u,s$. Right panel: valence-number sum rule ratio $R_{j_{1v},j_2}^N(x_2,\mu_1^2,\mu_2^2)$ defined in Eq.~\eqref{eq_numSumRule}, shown for the indicated valence channels. Lines with filled markers show the production configuration, with $N_x^{\mathrm{solver}}=90$, $N_x^{\mathrm{stored}}=56$, and $N_t=41$, whereas open markers without connecting lines show the refined configuration, with $N_x^{\mathrm{solver}}=150$, $N_x^{\mathrm{stored}}=70$, and $N_t=60$. The calculation is performed for $\mu_1^2=100~\mathrm{GeV}^2$ and $\mu_2^2=25~\mathrm{GeV}^2$. Exact preservation of the corresponding sum rule gives a ratio equal to unity.
}
	\label{fig:unequal_sum_rules}
\end{figure}

The left panel of Fig.~\ref{fig:unequal_sum_rules} shows $R_j^M$ for $\mu_1^2=100~\mathrm{GeV}^2$ and $\mu_2^2=25~\mathrm{GeV}^2$, for fixed second parton $j=g,u,s$. The ratio remains close to unity over the full $x_2$ range shown. This confirms that the unequal-scale single-leg evolution implemented in \texttt{ChromaPDFEvolver} preserves the DPDF momentum sum rule to good numerical accuracy. The small residual deviations from unity are attributed to finite grid resolution and numerical integration effects.

We next investigate the valence-number sum rule in the same way, by comparing the unequal-scale number integral with the corresponding equal-scale DPDF number integral at the lower scale. Analogously to the momentum case, we define
\begin{equation}
	\label{eq_numSumRule}
	R_{j_{1v},j_2}^N(x_2,\mu_1^2,\mu_2^2)
	=
	\frac{
		\displaystyle
		\int_0^{1-x_2} dx_1
		D_{j_{1v} j_2}(x_1,x_2,\mu_1^2,\mu_2^2)
	}{
		\displaystyle
		\int_0^{1-x_2} dx_1
		D_{j_{1v} j_2}(x_1,x_2,\mu_2^2,\mu_2^2)
	}.
\end{equation}
If the lower scale equal-scale DPDF satisfies the Gaunt-Stirling number sum rule, the denominator is equal to the appropriate coefficient in Eq.~\eqref{dpdf_number_sum_rule} multiplied by $f_{j_2}(x_2,\mu_2^2)$. Hence, Eq.~\eqref{eq_numSumRule} tests whether the single-leg evolution preserves the valence-number constraint.

The numerical results for Eq.~\eqref{eq_numSumRule} are shown in the right panel of Fig.~\ref{fig:unequal_sum_rules}. The number sum rule ratios remain close to unity, although the $u_v u$ and $u_v\bar u$ channels show percent-level deviations, with the largest deviation occurring for $u_v u$. The channel dependence observed in Fig.~\ref{fig:unequal_sum_rules} can be understood as a channel-dependent numerical sensitivity of the number sum rule test. In the exact single-leg evolution, the integral over the first parton valence combination is conserved, and hence all ratios in Eq.~\eqref{eq_numSumRule} should be equal to unity. However, the numerical residual depends on the structure of the particular DPDF combination. In the \texttt{GS09} approach~\cite{Gaunt:2009re}, equal flavor quark-antiquark correlations enter through the non-factorized $j\bar{j}$ contribution associated with the sPDF feed term, while the $uu$ component also contains a valence-number correction. Consequently,
\begin{equation}
	D_{u_vu}=D_{uu}-D_{\bar{u}u},
	\qquad
	D_{u_v\bar{u}}=D_{u\bar{u}}-D_{\bar{u}\bar{u}},
\end{equation}
so that both combinations receive the $u\bar{u}$ correlation contribution with different relative signs, while the $u_vu$ channel also inherits the valence-number modification. Since these quantities are differences of individual flavor distributions, numerical errors can also be relatively amplified. The $u_vg$ channel does not involve the same equal flavor modifications, which explains why it is numerically more stable.

The deviations are largest at intermediate values of $x_2$. This is consistent with the discussion of Ref.~\cite{Gaunt:2009re}, where the effects of the valence-number and $j\bar j$ correlation terms were found to be most visible in the intermediate-$x$ region. At smaller $x$, these effects are diluted by sea-dominated contributions, while at larger $x$ phase space effects become dominant. This explains why the $u_vu$ and $u_v\bar u$ curves show their largest deviations in the intermediate-$x_2$ region and move back toward unity at both smaller and larger $x_2$. The remaining deviations are also consistent with the expected numerical accuracy of the interpolation and integration procedure.

To test this interpretation, we repeated the \texttt{GS09} calculation using the refined numerical configuration described in Sec.~\ref{sec:ddglap}, while keeping all other evolution settings and grid ranges unchanged. As shown in the right panel of Fig.~\ref{fig:unequal_sum_rules}, the maximum deviations of the $u_vu$ and $u_v\bar{u}$ ratios decrease from approximately $1.7\%$ to $1.4\%$ and from approximately $0.8\%$ to $0.5\%$, respectively, while the $u_vg$ ratio remains within approximately $0.2\%$ of unity for both configurations. This supports the conclusion that finite grid resolution contributes to the residual deviations without changing their channel-dependent pattern. Since the differences between the two configurations are small, whereas the refined configuration substantially increases the storage and computational requirements of the generated DPDF tables, we retain the production grid for the subsequent UDPDF calculations.

\begin{figure}[htbp]
	\centering
	\includegraphics[width=0.49\textwidth]{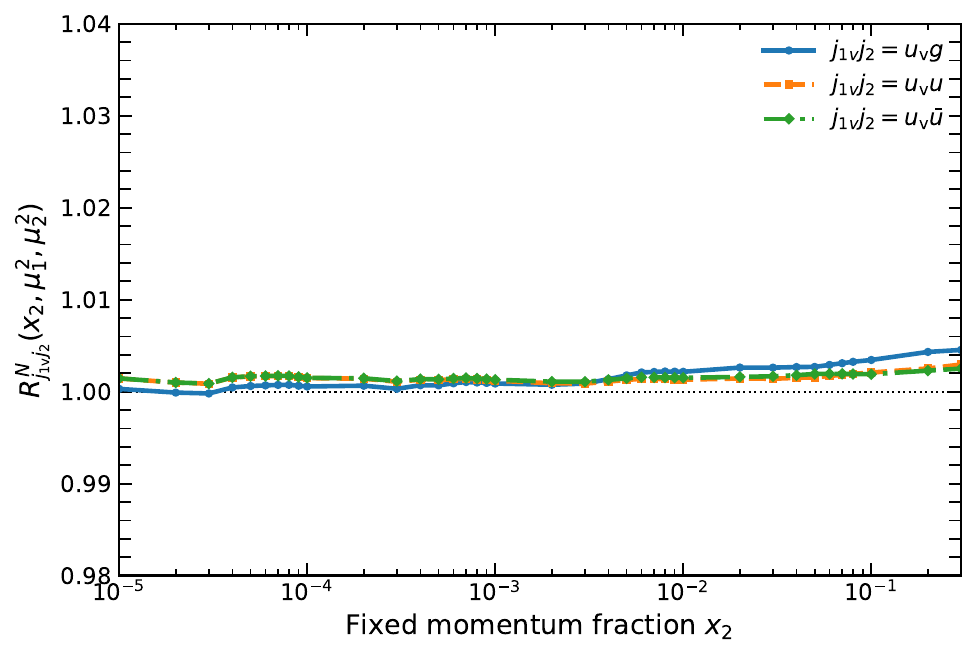}
	\includegraphics[width=0.49\textwidth]{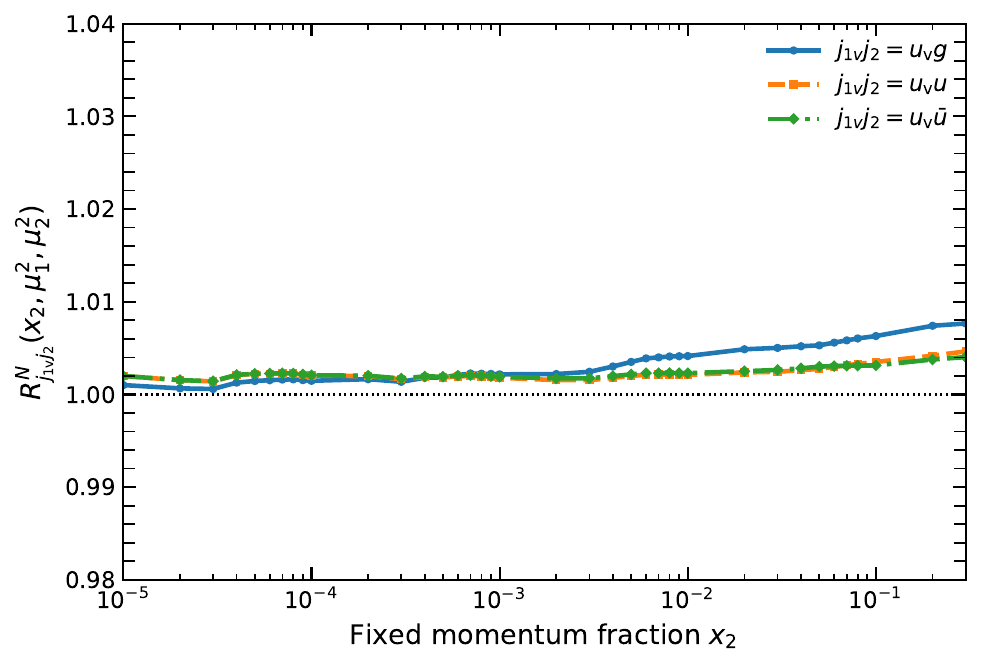}
\caption{
	Valence-number sum rule test obtained using the smooth factorized equal-scale input in Eq.~\eqref{eq:smooth_factorized_test}. The left and right panels correspond to the phase space exponents $n=1$ and $n=2$, respectively. Both calculations use the production configuration, with $N_x^{\mathrm{solver}}=90$, $N_x^{\mathrm{stored}}=56$, and $N_t=41$.
}
	\label{fig:factorized_number_sum_rules}
\end{figure}

We also replaced the \texttt{GS09} equal-scale input by the smoother factorized distributions
\begin{equation}
	\label{eq:smooth_factorized_test}
	D_{ij}^{(n)}(x_1,x_2,\mu_0^2)
	=
	f_i(x_1,\mu_0^2)
	f_j(x_2,\mu_0^2)
	(1-x_1-x_2)^n
	\Theta(1-x_1-x_2),
	\qquad n=1,2.
\end{equation}
Using the production configuration, the pronounced opposite-sign deviations of the $u_vu$ and $u_v\bar{u}$ channels obtained with the \texttt{GS09} input are no longer observed for either $n=1$ or $n=2$, and the three channels display a much more similar behavior over most of the shown $x_2$ range; see Fig.~\ref{fig:factorized_number_sum_rules}. Together with the refined-grid result, this supports the interpretation that the observed deviations arise from the interplay between finite numerical resolution and the flavor-dependent structure of the \texttt{GS09} input, rather than from a general instability of the unequal-scale single-leg evolution.

	We next test whether the different UDPDF approaches reproduce the corresponding collinear DPDF after integration over the transverse momenta. This is an important check because a sizeable normalization mismatch can obscure the interpretation of model comparisons. For the DMKMRW model the normalization follows directly from Eq.~\eqref{eq:MKMRW_UDPDF_norm}, since the transverse kernels are normalized by construction. For the DVO-MRW model, Eq.~\eqref{eq:LOMRW_norm} provides the corresponding normalization criterion against which we test the numerical construction.
	
\begin{figure}[htbp]
		\centering
		
		\begin{subfigure}{0.48\textwidth}
			\centering
			\includegraphics[width=\textwidth]{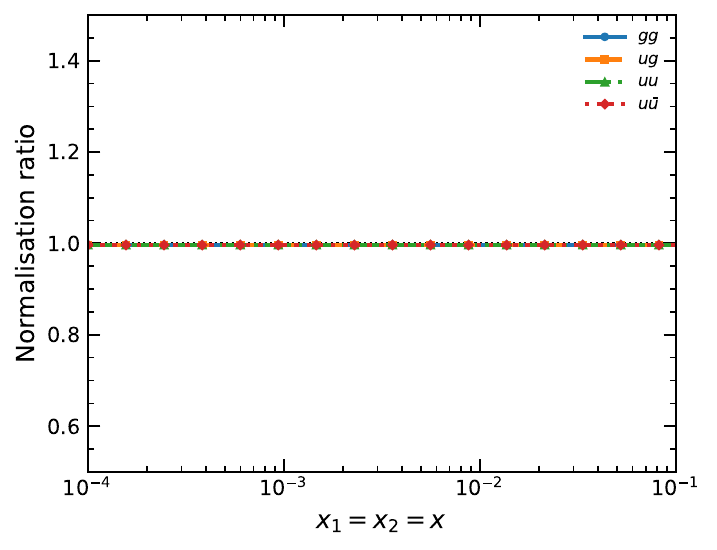}
			\caption{DMKMRW}
			\label{fig:normalization_dmkmrw}
		\end{subfigure}
		\hfill
		\begin{subfigure}{0.48\textwidth}
			\centering
			\includegraphics[width=\textwidth]{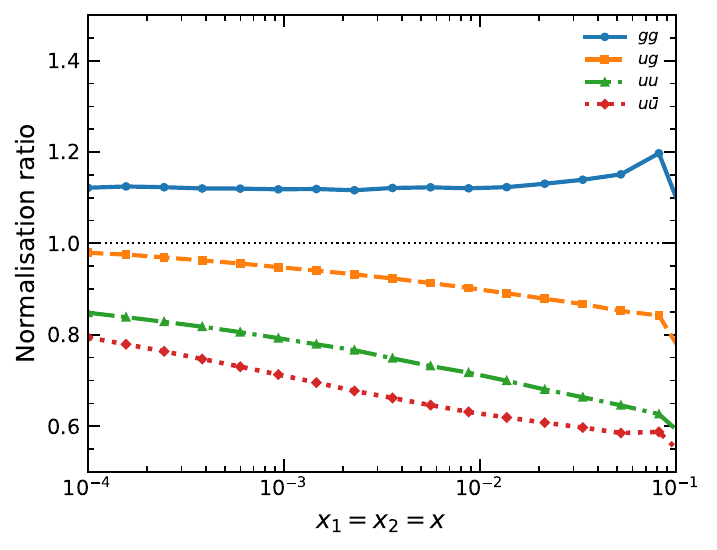}
			\caption{DVO-MRW}
			\label{fig:normalization_dvo_mrw}
		\end{subfigure}
		
		\caption{
			Normalization test for the UDPDFs. The plotted quantity is the ratio of the transverse momentum integrated UDPDF to the corresponding collinear DPDF,
			$\int dk_{1t}^2 dk_{2t}^2\,\mathcal{F}_{ij}/D_{ij}$.
			The calculation is performed for $\mu_1^2=\mu_2^2=100~\mathrm{GeV}^2$. Results are shown for the $uu$, $u\bar u$, $ug$, and $gg$ channels. Exact normalization corresponds to unity.
		}
		\label{fig:normalization}
	\end{figure}

	The results are shown in Fig.~\ref{fig:normalization}. The DMKMRW construction reproduces the input collinear DPDF with good numerical accuracy for all channels shown. This is expected from the normalized-kernel structure of the model. In contrast, the DVO-MRW construction is not exactly normalization preserving. For the quark-containing channels the ratio is close to unity only at small $x$, while larger deviations appear as $x$ increases. The gluon-gluon channel shows the largest normalization excess, indicating that the virtuality ordered last step prescription can generate an integrated distribution larger than the input collinear DPDF.
	
	This behavior motivates the normalization matched DVO-MRW construction introduced above. The purpose of MDVO-MRW is not to change the virtuality ordered transverse momentum shape, but to remove the overall normalization mismatch by a multiplicative matching factor. In this way, comparisons involving MDVO-MRW are less affected by integrated normalization differences and more directly reflect differences in the generated $k_t$-dependence. In addition, because the DLO-MRW implementation used in this work is restricted to the perturbative region $k_t \geq \mu_0$, we do not include a normalization ratio test for this prescription. However, as discussed in Ref.~\cite{Guiot:2022psv}, the conventional LO-MRW model can suffer from normalization ambiguities, which is one of the motivations for the MKMRW model.
	
We now compare factorized and non-factorized UDPDFs as functions of
$k_t^2=k_{1t}^2=k_{2t}^2$
at $\mu_1=\mu_2=\mu=100~\mathrm{GeV}$, with $x_1=x_2=x$. The factorized approximation is obtained by multiplying the corresponding single UPDFs, whereas the non-factorized distributions are constructed from the full \texttt{GS09} DPDFs.

\begin{figure}[htbp]
		\centering
		\begin{subfigure}{0.48\textwidth}
			\centering
			\includegraphics[width=\textwidth]{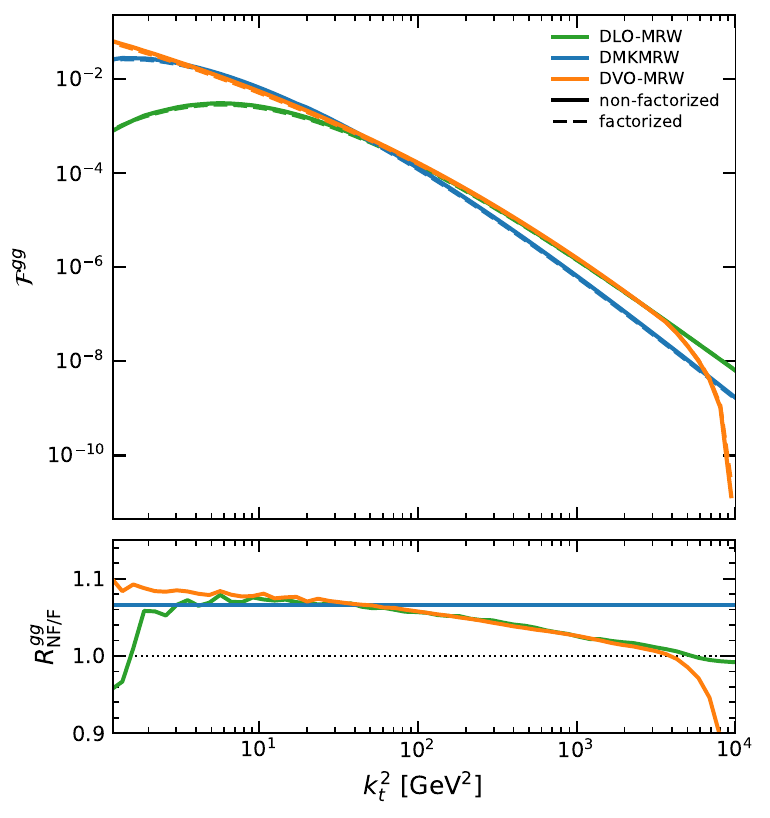}
			\caption{\(gg\)}
		\end{subfigure}
		\begin{subfigure}{0.48\textwidth}
			\centering
			\includegraphics[width=\textwidth]{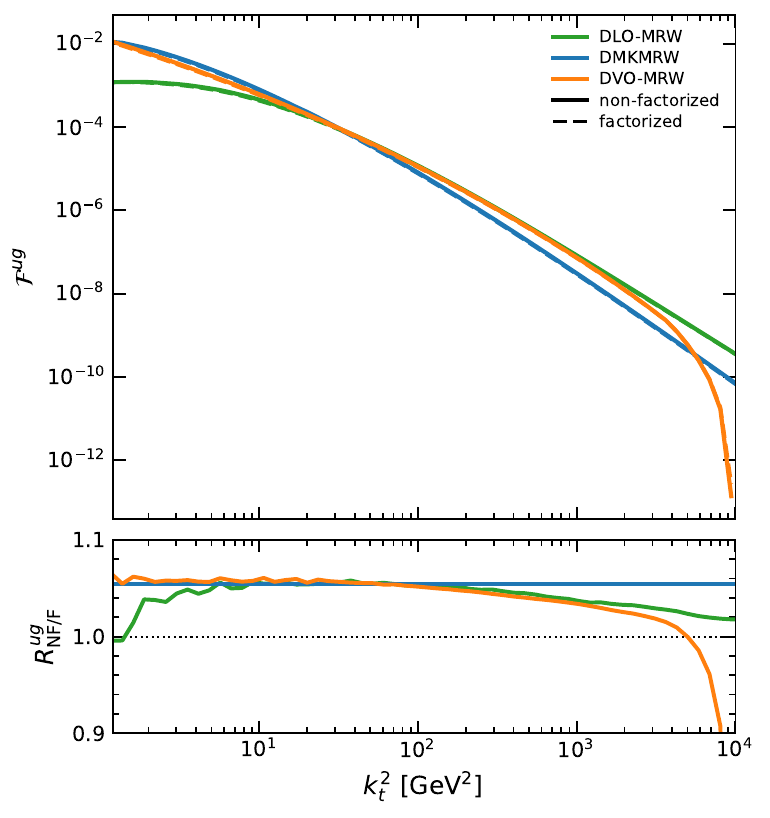}
			\caption{\(ug\)}
		\end{subfigure}
		
		\begin{subfigure}{0.48\textwidth}
			\centering
			\includegraphics[width=\textwidth]{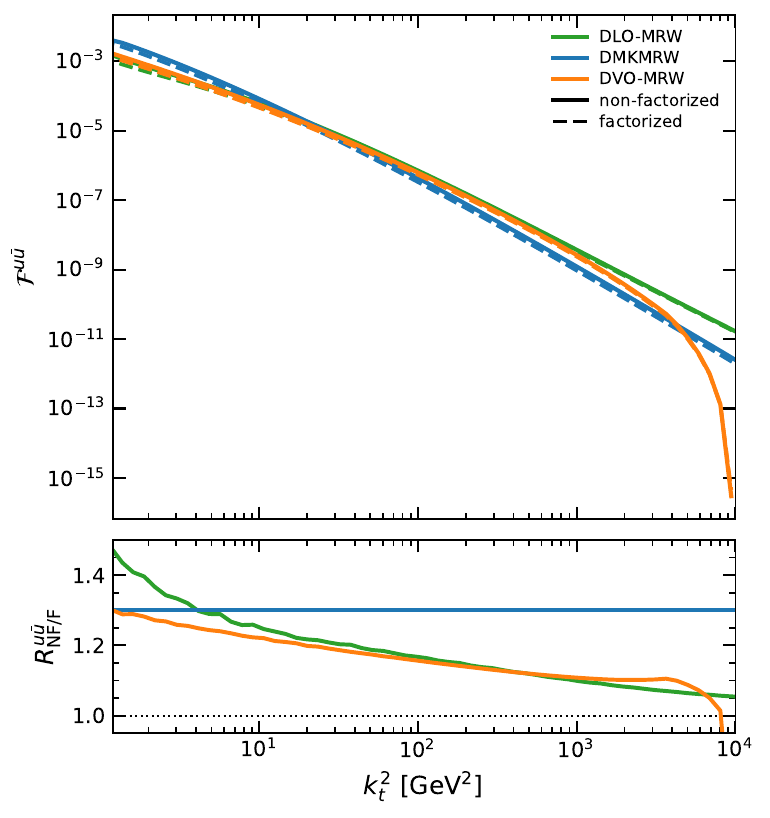}
			\caption{\(u\bar u\)}
		\end{subfigure}
		\begin{subfigure}{0.48\textwidth}
			\centering
			\includegraphics[width=\textwidth]{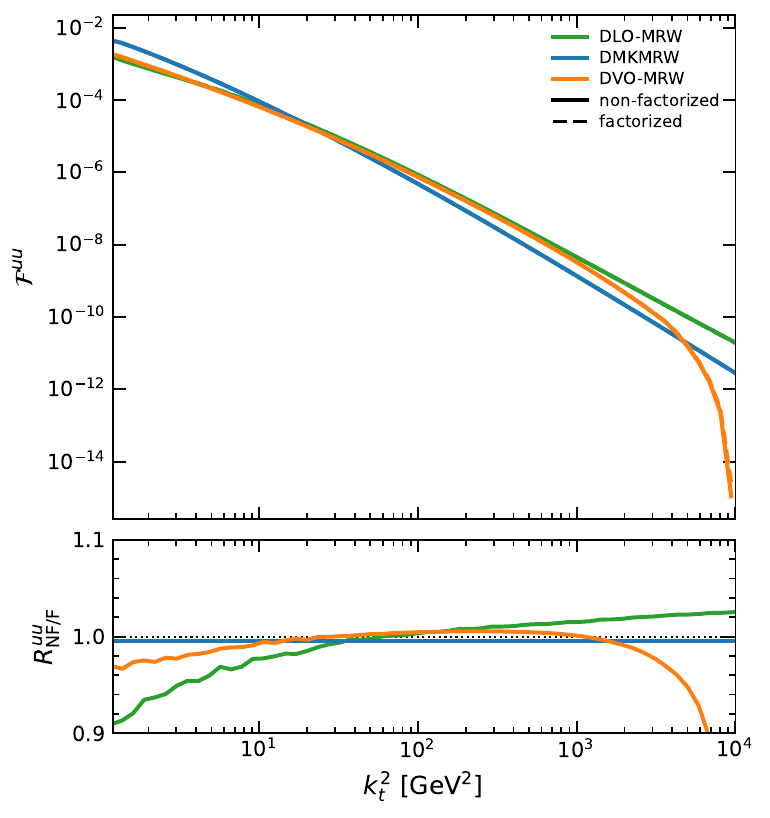}
			\caption{\(uu\)}
		\end{subfigure}
		\caption{
			Representative $k_t^2$-dependence of the UDPDFs at $x_1=x_2=0.01$ and $\mu_1=\mu_2=100~\mathrm{GeV}$. The upper panel in each subfigure shows the factorized and non-factorized results, while the lower panel shows the ratio of the non-factorized distribution to the corresponding factorized approximation.
		}
		\label{fig:x01DUDPDF}
	\end{figure}

Representative $k_t^2$-dependent distributions are shown in Fig.~\ref{fig:x01DUDPDF} for $x_1=x_2=0.01$. The upper panels show the absolute UDPDFs, while the lower panels show the ratio of the non-factorized distribution to the corresponding factorized approximation. This ratio measures the size of longitudinal double parton correlations after the transverse momentum dependence has been generated.

A simple but important difference between the prescriptions is visible. In the DMKMRW construction, the transverse dependence factorizes from the collinear DPDF,
\begin{equation}
	\label{eq:DMKMRW_shape_factorization_results}
	\mathcal F_{ij}^{\mathrm{DMKMRW}}
	=
	D_{ij}(x_1,x_2,\mu_1^2,\mu_2^2)
	\mathcal K_i^{\mathrm{MKMRW}}(k_{1t}^2,\mu_1^2)
	\mathcal K_j^{\mathrm{MKMRW}}(k_{2t}^2,\mu_2^2).
\end{equation}
Consequently, the ratio of the non-factorized DMKMRW distribution to the corresponding factorized product is
\begin{equation}
	\label{eq:DMKMRW_ratio_factorization}
	\frac{
		\mathcal F_{ij}^{\mathrm{DMKMRW}}
	}{
		\mathcal F_{ij}^{\mathrm{MKMRW,fact.}}
	}
	=
	\frac{
		D_{ij}(x_1,x_2,\mu_1^2,\mu_2^2)
	}{
		f_i(x_1,\mu_1^2)f_j(x_2,\mu_2^2)
	}.
\end{equation}
The transverse kernels cancel in this ratio. Therefore, in the DMKMRW model, non-factorized DPDF effects modify the normalization and flavor dependence of the UDPDF, but not the $k_t$-shape.

The direct DLO-MRW and virtuality ordered DVO-MRW constructions behave differently from the normalized-kernel DMKMRW model. In both cases the DPDF is evaluated inside the last step convolution, so the non-factorized DPDF can affect not only the normalization but also the generated $k_t$ dependence. In the DLO-MRW prescription, the DPDF entering the convolution is evaluated at the transverse scales $k_{1t}^2$ and $k_{2t}^2$. In the DVO-MRW model the corresponding scales are instead the virtualities
$$
K_1^2=\frac{k_{1t}^2}{1-z_1},
\qquad
K_2^2=\frac{k_{2t}^2}{1-z_2}.
$$
Thus, for fixed $k_t$, the DVO-MRW prescription samples the DPDF at larger scales than the direct DLO-MRW prescription.

This difference helps explain the small-$k_t^2$ behavior in Fig.~\ref{fig:x01DUDPDF}. Near the lower perturbative boundary, the DLO-MRW distribution samples the DPDF close to the input scale and is accompanied by a Sudakov factor over a relatively long evolution interval. In contrast, the DVO-MRW convolution samples the DPDF at the larger virtuality scale $K^2$ and with a shorter Sudakov evolution interval. This tends to make the DVO-MRW distribution larger than the perturbative DLO-MRW result at small $k_t^2$, especially in gluon dominated channels.

At larger $k_t$, the virtuality ordering condition $K^2<\mu^2$ restricts the allowed $z$ region in the DVO-MRW model. This pushes the parent fractions $x_1/z_1$ and $x_2/z_2$ closer to the kinematic boundary,
$$
\frac{x_1}{z_1}+\frac{x_2}{z_2}=1,
$$
where the \texttt{GS09} DPDFs are strongly constrained by momentum and number sum rules and by phase space suppression. As a result, the DVO-MRW curves decrease rapidly as the virtuality ordered phase space closes. By contrast, the angular ordered DLO-MRW prescription is less restrictive in this region and can produce a comparatively larger large $k_t$ tail. This $k_t$-dependent suppression mechanism is absent in the DMKMRW model, where the transverse dependence is generated by external normalized kernels rather than by a last step convolution with the DPDF.

The channel dependence is also significant. The $u\bar u$ channel can show an enhancement of the non-factorized distribution relative to the factorized approximation, reflecting quark-antiquark correlations in the \texttt{GS09} input~\cite{Gaunt:2009re}. This enhancement is especially visible in the DLO-MRW ratio, where the DPDF is sampled more directly at the scale $k_t^2$. In contrast, the $uu$ channel is more strongly affected by valence number constraints, especially at larger $x$. The gluon dominated channels, $gg$ and $ug$, show more moderate deviations from unity.

Generally, Fig.~\ref{fig:x01DUDPDF} shows that the DMKMRW, DLO-MRW, and DVO-MRW prescriptions generate visibly different transverse momentum behavior. In the DMKMRW model, the large-$k_t$ behavior is controlled by the external normalized kernel. In the DLO-MRW model, it is controlled by the angular ordered last step construction. In the DVO-MRW model, it is strongly restricted by the virtuality ordering condition. The large $k_t$ region is particularly important because large transverse momentum tails in conventional MRW type UPDFs have already been discussed in the literature and can affect differential cross section predictions in regions sensitive to large incoming partonic transverse momenta~\cite{Guiot:2022psv,Valeshabadi:2021twn,Valeshabadi:2021spa}. Nevertheless, the implications for physical observables cannot be inferred from the UDPDFs alone. They depend on the off-shell hard matrix element, the phase space cuts, and the measured observable. A dedicated phenomenological analysis of DPS final states is therefore required to quantify the numerical impact of the differences observed here.
	
		To study the correlation effects over a wider kinematic range, we show contour maps in the $(x,k_t)$ plane with
	$$
	x_1=x_2=x,
	\qquad
	k_{1t}=k_{2t}=k_t,
	\qquad
	\mu^2=10^4~\mathrm{GeV}^2.
	$$
	The color scale represents $\log_{10}R(x,k_t)$, where $R$ denotes the ratio specified in each figure. Thus, $\log_{10}R=0$ corresponds to equality between the numerator and denominator. The black contour in the figures indicates $R=1$. Regions with $\log_{10}R>0$ correspond to enhancement of the numerator, while regions with $\log_{10}R<0$ correspond to suppression.
	
	We do not display a separate DMKMRW non-factorized to factorized contour map in the text, because Eq.~\eqref{eq:DMKMRW_ratio_factorization} already shows that this ratio is independent of $k_t$. The corresponding result therefore mainly probes the collinear longitudinal correlation factor $D_{ij}/(f_i f_j)$. We also do not display separate DLO-MRW contour maps, because the DLO-MRW prescription is already compared with the other prescriptions in the fixed-$x$ $k_t^2$ spectra of Fig.~\ref{fig:x01DUDPDF}, which directly shows its perturbative region behavior. As representative examples, we therefore limit the non-factorized to factorized contour maps to those for the DVO-MRW and MDVO-MRW prescriptions, where the non-factorized DPDF enters inside the virtuality-ordered last-step convolution and can change the transverse momentum dependence itself. Additionally, we do not repeat all model to model ratios as contour maps. Instead, the contour plots are used to show the two-dimensional structure of the non-factorized corrections, together with one representative model-dependence comparison between the two normalization-preserving prescriptions.
	
\begin{figure}[htbp]
	\centering
	\begin{subfigure}{0.48\textwidth}
		\centering
		\includegraphics[width=\textwidth]{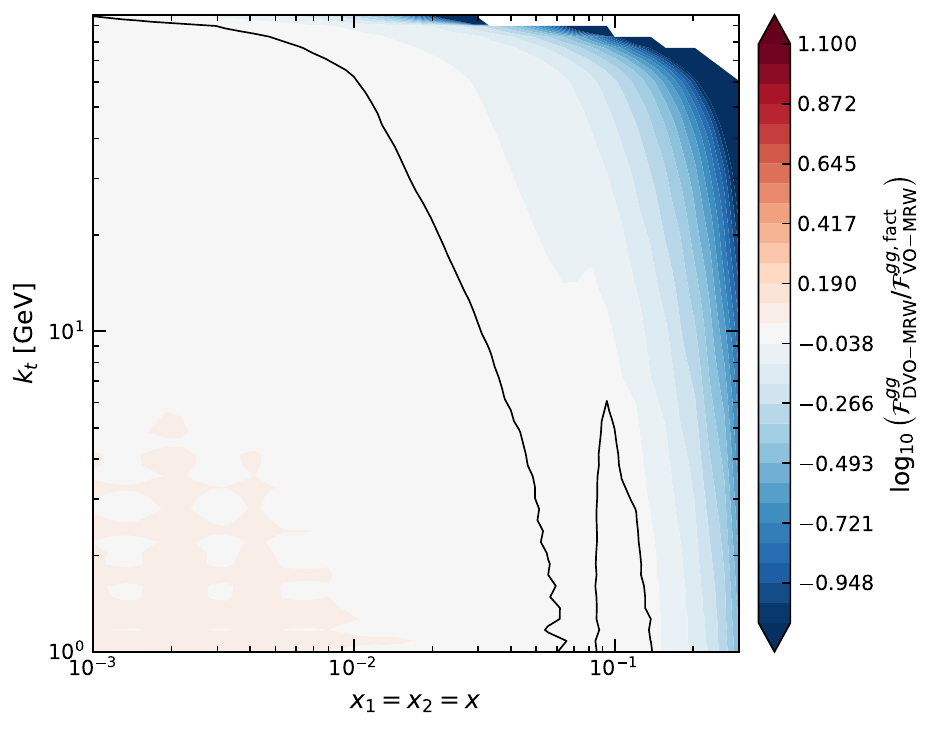}
		\caption{\(gg\)}
	\end{subfigure}
	\hfill
	\begin{subfigure}{0.48\textwidth}
		\centering
		\includegraphics[width=\textwidth]{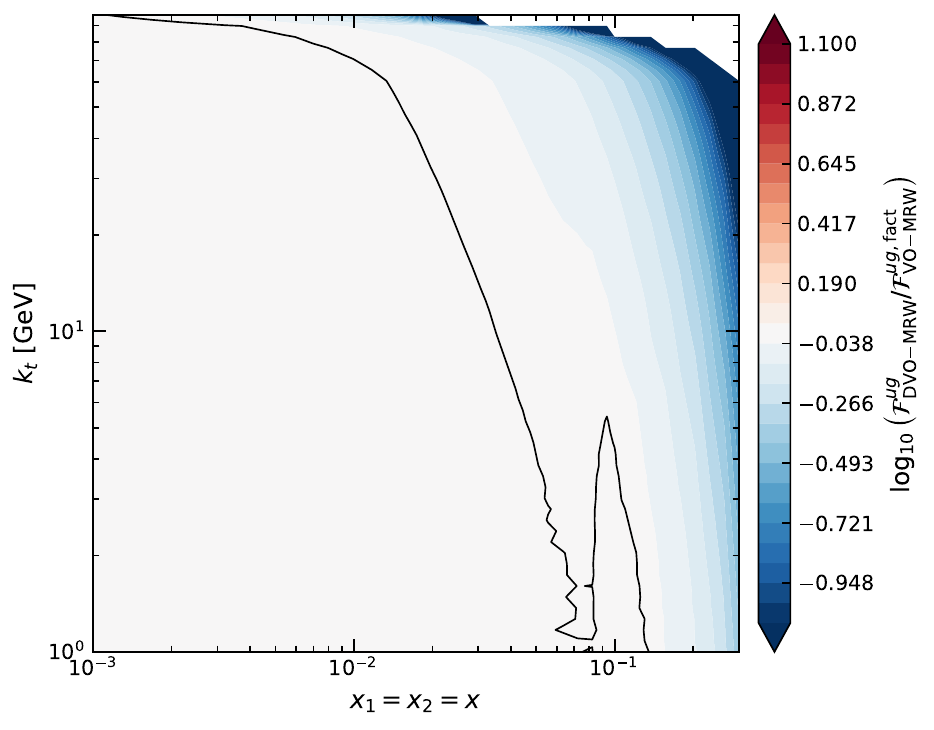}
		\caption{\(ug\)}
	\end{subfigure}
	
	\vspace{0.15cm}
	
	\begin{subfigure}{0.48\textwidth}
		\centering
		\includegraphics[width=\textwidth]{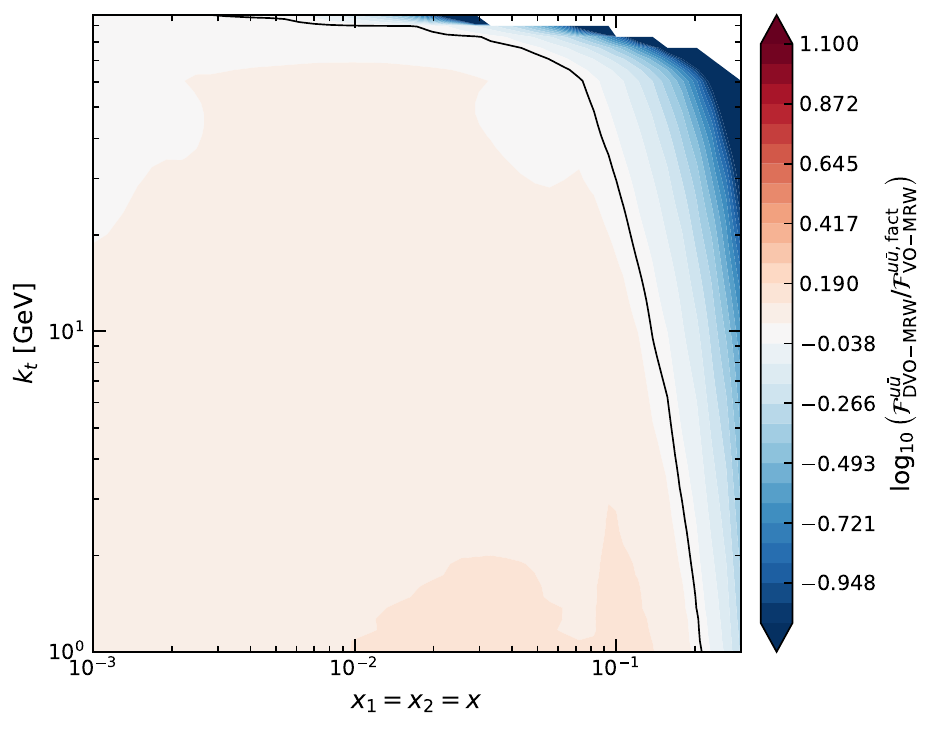}
		\caption{\(u\bar u\)}
	\end{subfigure}
	\hfill
	\begin{subfigure}{0.48\textwidth}
		\centering
		\includegraphics[width=\textwidth]{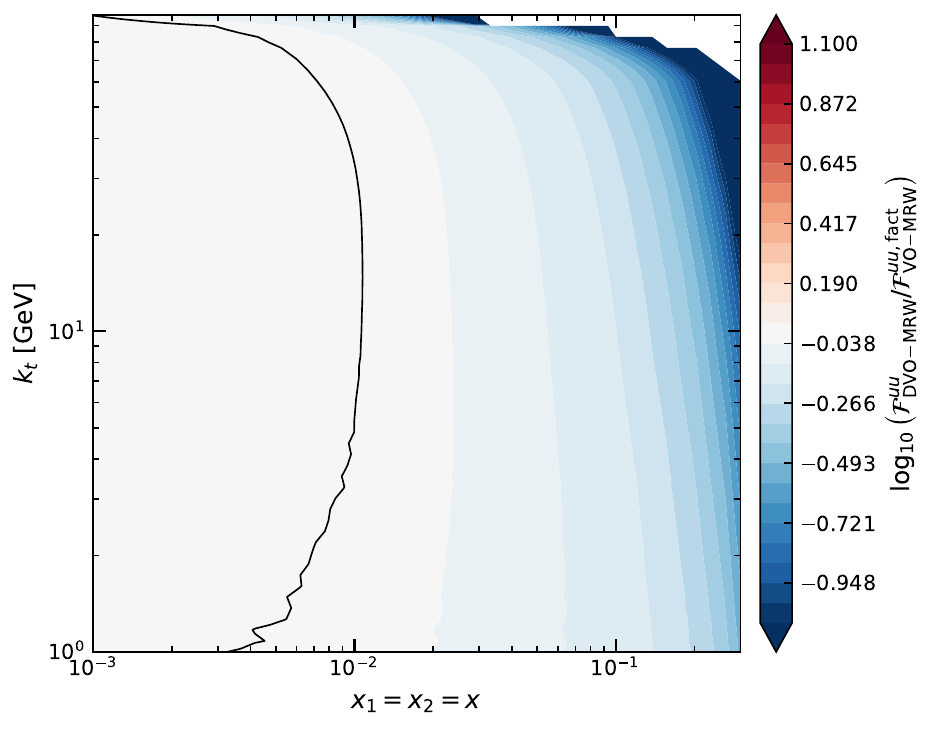}
		\caption{\(uu\)}
	\end{subfigure}
	
	\caption{
		Contour maps of the non-factorized to factorized ratio for the virtuality ordered MRW construction, $R_{ij}^{\mathrm{DVO-MRW}}=\mathcal{F}_{ij}^{\mathrm{DVO-MRW}}/\mathcal{F}_{ij}^{\mathrm{VO-MRW,fact.}}$, for $x_1=x_2=x$, $k_{1t}=k_{2t}=k_t$, and $\mu^2=10^4~\mathrm{GeV}^2$. The color scale represents $\log_{10}R_{ij}^{\mathrm{DVO-MRW}}$, while the black contour indicates $R_{ij}^{\mathrm{DVO-MRW}}=1$.
	}
	\label{fig:vo_nonfact_over_fact_contours}
\end{figure}

	Figure~\ref{fig:vo_nonfact_over_fact_contours} shows the ratio
	\begin{equation}
		\label{eq:DVO_factorized_ratio}
		R_{ij}^{\mathrm{DVO}}(x,k_t)
		=
		\frac{
			\mathcal F_{ij}^{\mathrm{DVO-MRW}}
		}{
			\mathcal F_{ij}^{\mathrm{VO-MRW,fact.}}
		}.
	\end{equation}
	This figure demonstrates that non-factorized effects in the DVO-MRW model are not simply overall normalization corrections. They depend on both $x$ and $k_t$, with the strongest suppression occurring at large $x$ and large $k_t$. As it is discussed, this is the region where the virtuality ordered convolution samples the DPDF closest to the boundary $x_1+x_2=1$. The effect is therefore most relevant for DPS observables that probe moderate or large parton momentum fractions or large incoming transverse momenta.
	
\begin{figure}[htbp]
	\centering
	\begin{subfigure}{0.48\textwidth}
		\centering
		\includegraphics[width=\textwidth]{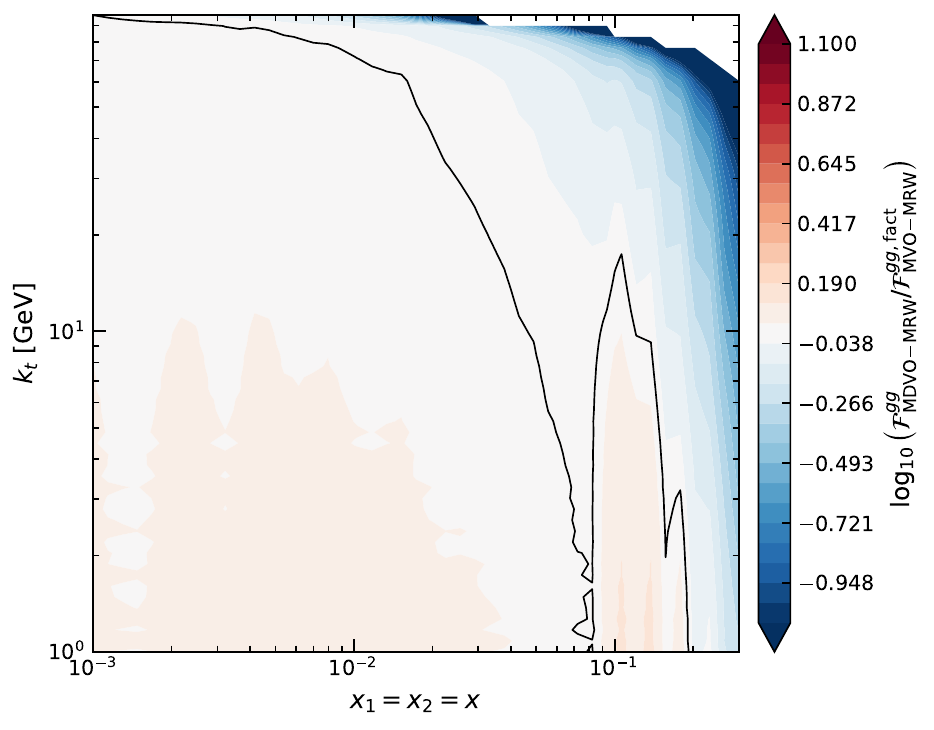}
		\caption{\(gg\)}
	\end{subfigure}
	\hfill
	\begin{subfigure}{0.48\textwidth}
		\centering
		\includegraphics[width=\textwidth]{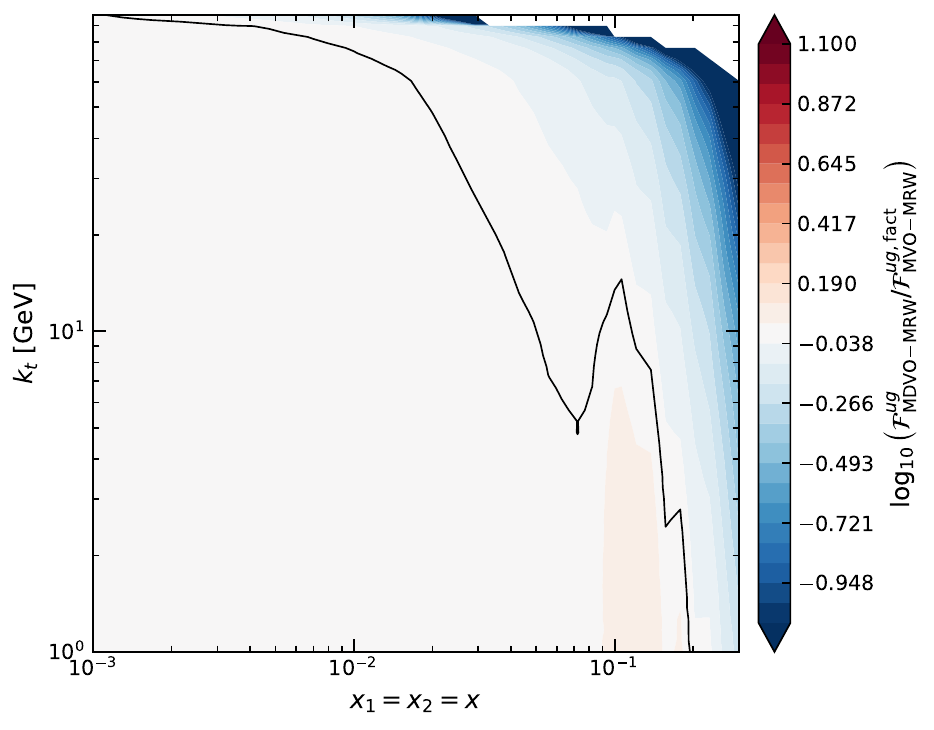}
		\caption{\(ug\)}
	\end{subfigure}
	
	\vspace{0.15cm}
	
	\begin{subfigure}{0.48\textwidth}
		\centering
		\includegraphics[width=\textwidth]{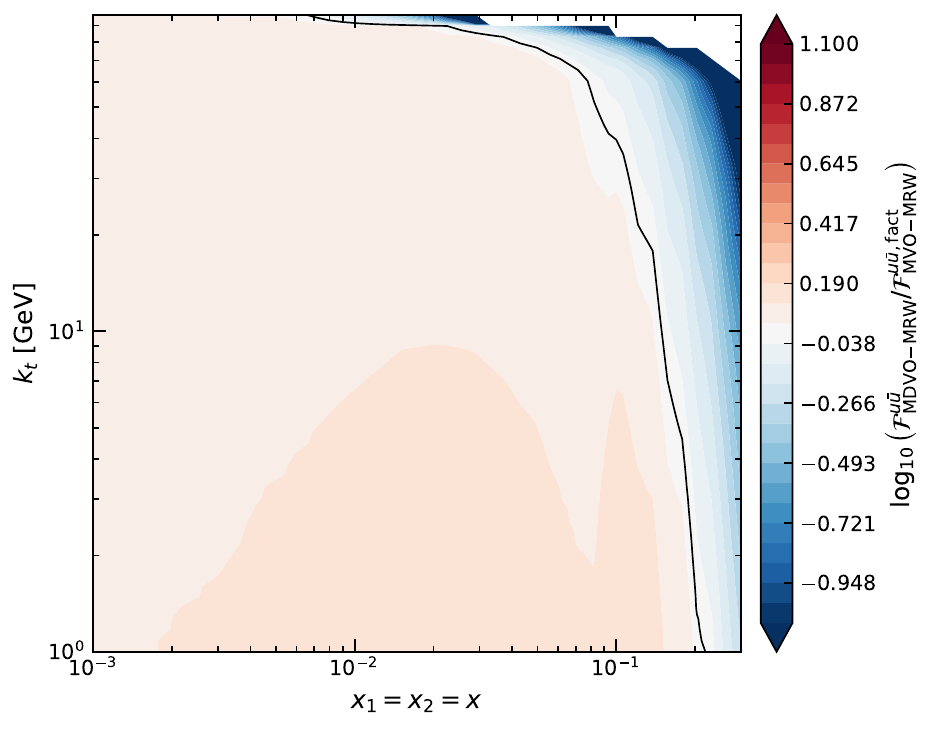}
		\caption{\(u\bar u\)}
	\end{subfigure}
	\hfill
	\begin{subfigure}{0.48\textwidth}
		\centering
		\includegraphics[width=\textwidth]{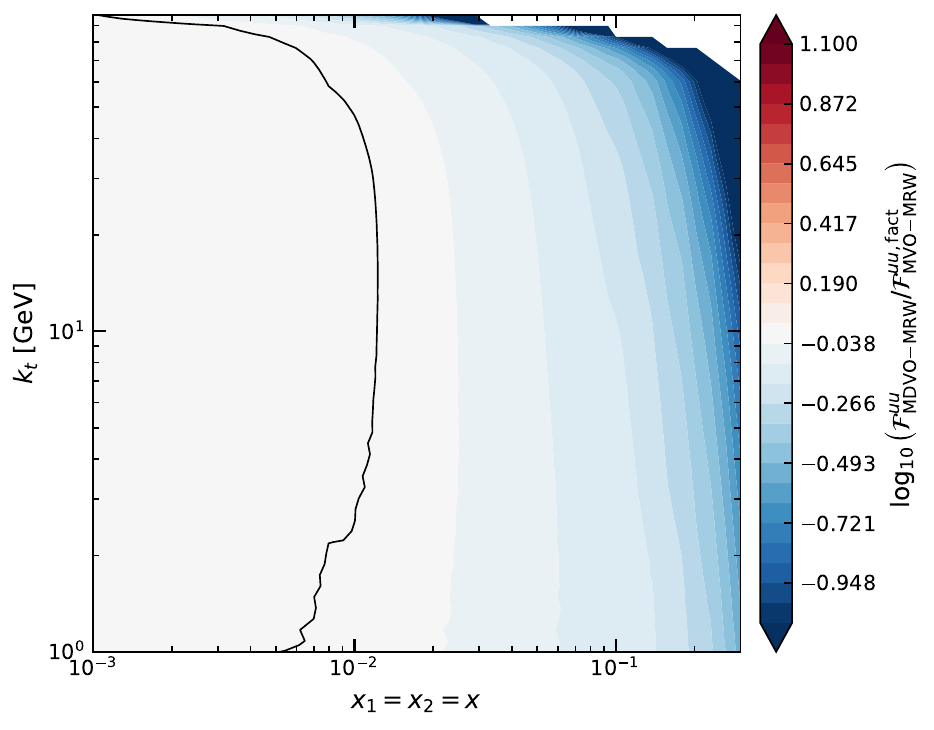}
		\caption{\(uu\)}
	\end{subfigure}
	
	\caption{
		Contour maps of the non-factorized to factorized ratio for the matched virtuality ordered MRW construction, $R_{ij}^{\mathrm{MDVO-MRW}}=\mathcal{F}_{ij}^{\mathrm{MDVO-MRW}}/\mathcal{F}_{ij}^{\mathrm{MVO-MRW,fact.}}$, with $x_1=x_2=x$, $k_{1t}=k_{2t}=k_t$, and $\mu^2=10^4~\mathrm{GeV}^2$. The color scale shows $\log_{10}R_{ij}^{\mathrm{MDVO-MRW}}$, and the black contour marks $R_{ij}^{\mathrm{MDVO-MRW}}=1$.
	}
	\label{fig:mvo_nonfact_over_fact_contours}
\end{figure}

	The corresponding matched result is shown in Fig.~\ref{fig:mvo_nonfact_over_fact_contours}, where
	\begin{equation}
		\label{eq:MDVO_factorized_ratio}
		R_{ij}^{\mathrm{MDVO-MRW}}(x,k_t)
		=
		\frac{
			\mathcal F_{ij}^{\mathrm{MDVO-MRW}}
		}{
			\mathcal F_{ij}^{\mathrm{MVO-MRW,fact.}}
		}.
	\end{equation}
	The matched version preserves the virtuality ordered transverse momentum shape while imposing the correct collinear normalization. The fact that the matched ratio still shows sizeable deviations from unity is important. It shows that the observed non-factorized effects are not merely artifacts of the integrated normalization mismatch of the DVO-MRW prescription. Instead, they reflect genuine longitudinal DPDF correlations sampled through the virtuality ordered last step convolution.
	
	A clear channel dependence is visible in both Figs.~\ref{fig:vo_nonfact_over_fact_contours} and \ref{fig:mvo_nonfact_over_fact_contours}. The suppression of the non-factorized distribution is strongest in the $uu$ channel. This behavior is expected from the DPDF number sum rule. Once one valence $u$ quark has been selected, the probability of finding a second $u$ quark is reduced compared with the naive product of two SPDFs. Therefore the factorized approximation tends to overestimate the $uu$ distribution, especially at larger $x$ and near the kinematic boundary.
	
	The opposite behavior can occur in the $u\bar u$ channel. In this case the non-factorized DPDF may become larger than the factorized product. This enhancement is related to the quark-antiquark correlation term already present in the input collinear DPDF at the initial scale~\cite{Gaunt:2009re}. After evolution and the virtuality ordered last step convolution, this correlation is reflected in the UDPDF ratio and can produce regions where
	$$
	\frac{\mathcal F_{u\bar u}^{\mathrm{non\mbox{-}fact.}}}
	{\mathcal F_{u\bar u}^{\mathrm{fact.}}}
	>1 .
	$$
	
\begin{figure}[htbp]
	\centering
	\begin{subfigure}{0.48\textwidth}
		\centering
		\includegraphics[width=\textwidth]{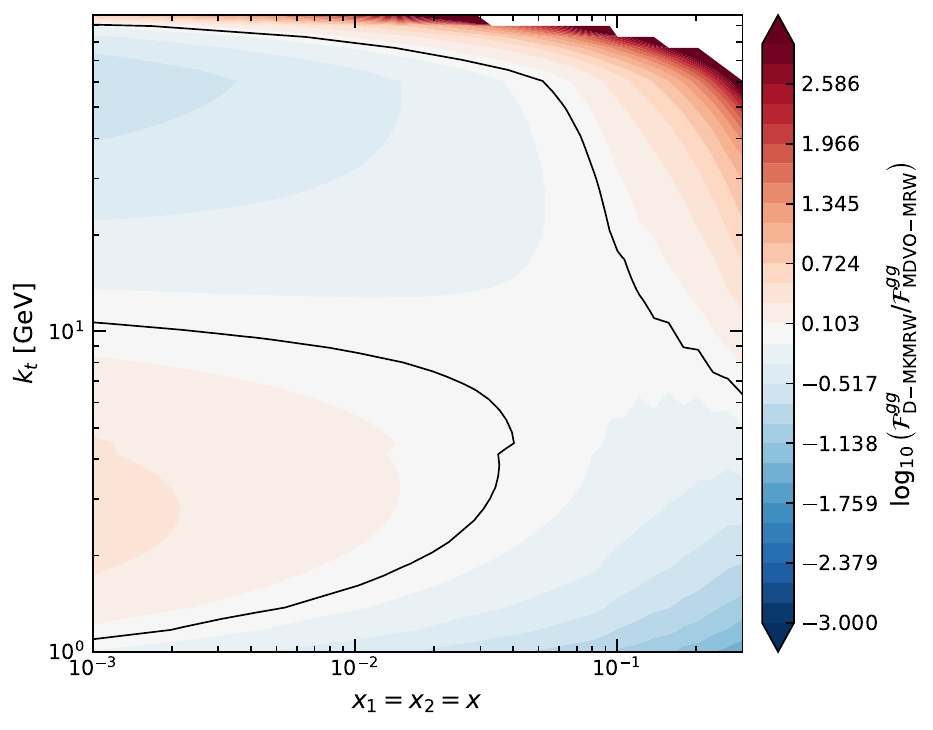}
		\caption{\(gg\)}
	\end{subfigure}
	\hfill
	\begin{subfigure}{0.48\textwidth}
		\centering
		\includegraphics[width=\textwidth]{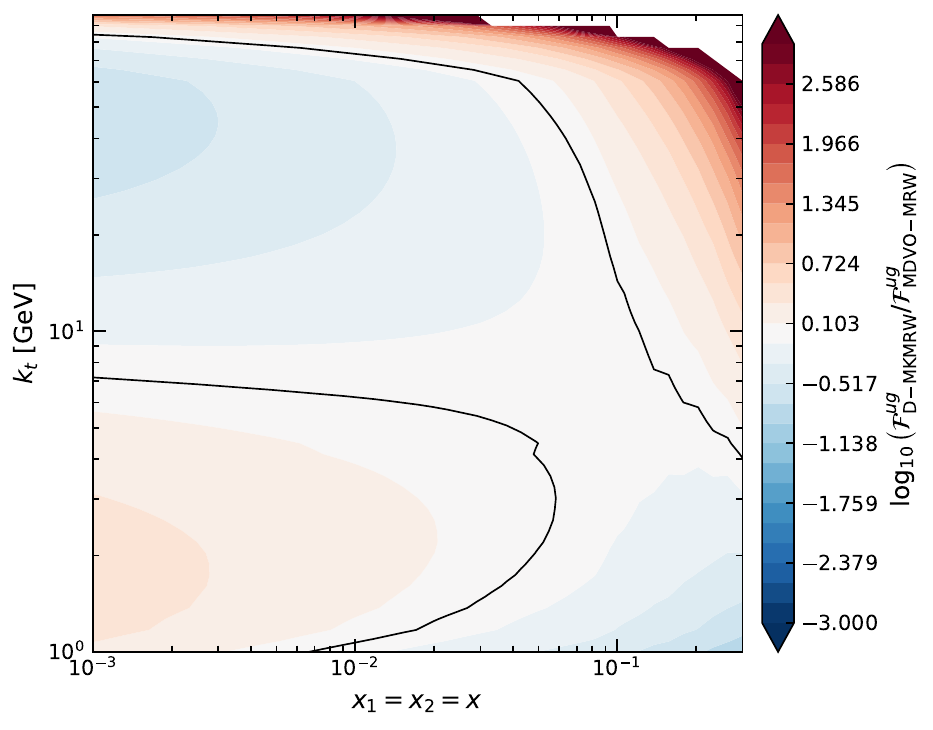}
		\caption{\(ug\)}
	\end{subfigure}
	
	\vspace{0.15cm}
	
	\begin{subfigure}{0.48\textwidth}
		\centering
		\includegraphics[width=\textwidth]{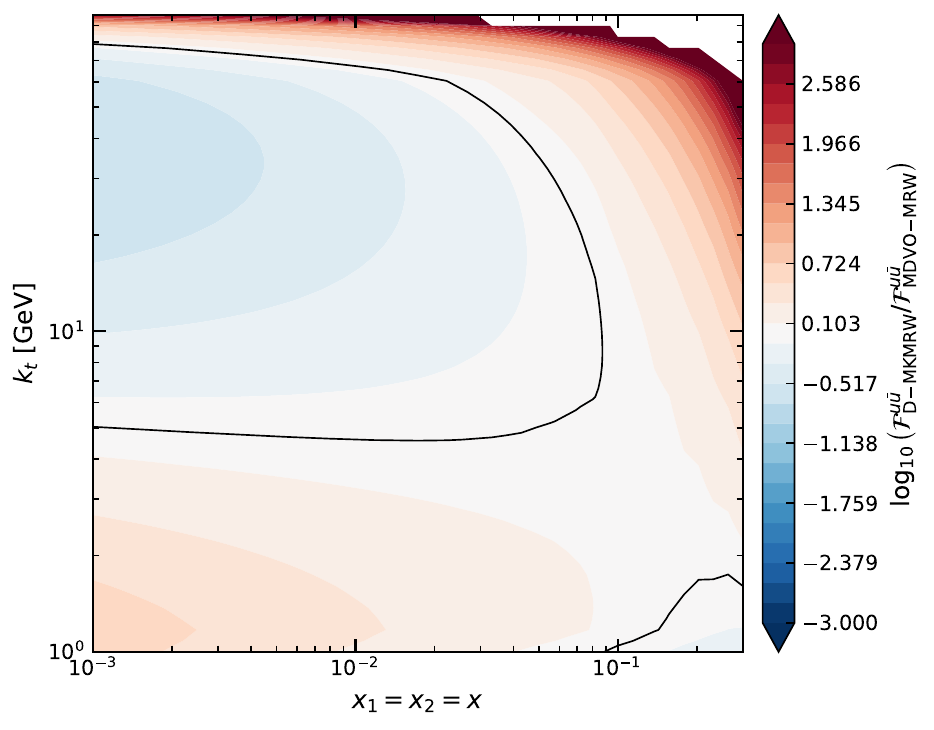}
		\caption{\(u\bar u\)}
	\end{subfigure}
	\hfill
	\begin{subfigure}{0.48\textwidth}
		\centering
		\includegraphics[width=\textwidth]{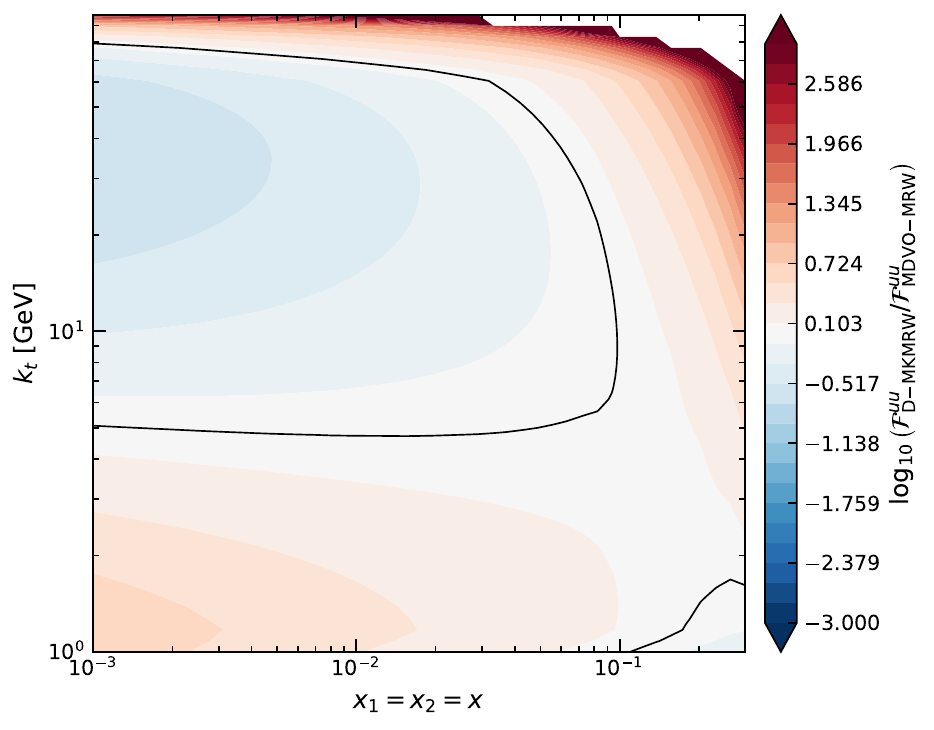}
		\caption{\(uu\)}
	\end{subfigure}
	
	\caption{
		Contour maps of the ratio $R_{ij}^{\mathrm{DMKMRW/MDVO-MRW}}=\mathcal{F}_{ij}^{\mathrm{DMKMRW}}/\mathcal{F}_{ij}^{\mathrm{MDVO-MRW}}$, for \(x_1=x_2=x\), $k_{1t}=k_{2t}=k_t$, and $\mu^2=10^4~\mathrm{GeV}^2$. The color scale shows $\log_{10}R_{ij}^{\mathrm{DMKMRW/MDVO-MRW}}$, while the black contour denotes $R_{ij}^{\mathrm{DMKMRW/MDVO-MRW}}=1$.
	}
	\label{fig:mkmrw_over_mvo_contours}
\end{figure}

	Finally, Fig.~\ref{fig:mkmrw_over_mvo_contours} compares two normalization-preserving UDPDFs through
	\begin{equation}
		\label{eq:model_dependence_ratio_matched}
		R_{ij}^{\mathrm{DMKMRW/MDVO-MRW}}(x,k_t)
		=
		\frac{
			\mathcal F_{ij}^{\mathrm{DMKMRW}}
		}{
			\mathcal F_{ij}^{\mathrm{MDVO-MRW}}
		}.
	\end{equation}
	The ratio varies strongly over the $(x,k_t)$ plane and depends on the partonic channel. This demonstrates that the difference between the normalized-kernel DMKMRW prescription and the matched virtuality ordered prescription is not a simple normalization effect. Both models reproduce the same collinear DPDF after integration over transverse momenta, but they generate different $k_t$ shapes. The model dependence is especially pronounced near the upper $k_t$ region, where the virtuality ordered model is strongly constrained by $K^2<\mu^2$, while the DMKMRW transverse dependence is generated by external normalized kernels.
	
	The comparison in Fig.~\ref{fig:mkmrw_over_mvo_contours} shows that the qualitative pattern observed in Fig.~\ref{fig:x01DUDPDF} remains present after the normalization matching is applied. Since the plotted quantity in Eq.~\eqref{eq:model_dependence_ratio_matched} is the ratio of the DMKMRW distribution to the MDVO-MRW distribution, regions with $R_{ij}^{\mathrm{DMKMRW/MDVO-MRW}}<1$ correspond to an enhancement of the matched virtuality ordered distribution relative to the DMKMRW one. In the gluon dominated channels, especially at small $k_t$, the ratio is below unity in part of the plotted region, indicating that the MDVO-MRW distribution is larger than the DMKMRW distribution there. In contrast, for the quark dominated channels in the same small $k_t$ region, the ratio is generally larger, showing that the DMKMRW distribution can be comparable to or larger than the MDVO-MRW result.
	
	At larger transverse momentum the virtuality ordering constraint in the MDVO-MRW approach becomes increasingly important. The virtuality ordering constraint $K^2<\mu^2$ reduces the available phase space as $k_t$ approaches the hard scale. Therefore, the MDVO-MRW distribution is suppressed relative to the DMKMRW distribution in the large $k_t$ region. This behavior is consistent with the direct $k_t^2$ spectra in Fig.~\ref{fig:x01DUDPDF}. The comparison therefore shows that the normalization matching does not remove the intrinsic shape difference between the normalized-kernel DMKMRW UDPDF and the virtuality ordered construction, and it only removes the overall normalization mismatch of the unmatched DVO-MRW model.
	
	The comparison identifies two separate sources of uncertainty in UDPDF phenomenology. The first is the longitudinal correlation contained in the input DPDF, which controls the difference between factorized and non-factorized distributions. The second is the prescription used to generate transverse momenta from the collinear input. In the DMKMRW model these two effects are largely separated, whereas in the DLO-MRW, DVO-MRW, and MDVO-MRW models they are coupled through the last step convolution. This coupling makes the non-factorized correction dependent not only on the input DPDF, but also on the scale choice, ordering variable, and phase-space restrictions of the last step construction. This distinction is important for future applications of UDPDFs to DPS observables, especially in regions where the process is sensitive to moderate or large parton momentum fractions or to large incoming transverse momenta.

	\section{Conclusion}
	\label{sec:conclusion}
	
	In this work we studied the construction of UDPDFs from non-factorized collinear DPDFs. The motivation is threefold. First, DPS calculations in a $k_t$-factorized framework require distributions in which the transverse momenta of both active partons are kept explicit. Second, introducing simple UDPDF model constructed from the full collinear DPDF is an essential tool for future $k_t$-factorization phenomenology. Third, the commonly used factorized approximation to DPDFs does not, in general, satisfy the double DGLAP evolution equations or the associated momentum and number sum rules. Therefore, it is important to understand how longitudinal DPDF correlations are reflected in the corresponding unintegrated distributions.
	
	As input we used the \texttt{GS09} DPDFs and evolved them to unequal scales with our numerical double DGLAP evolution framework, \texttt{ChromaPDFEvolver}. The unequal-scale evolution was validated by testing the momentum and valence-number sum rules. The numerical results show that the ratio defined in Eq.~\eqref{eq:unequal_scale_momentum_numerical_test} remains close to unity over the investigated range of $x_2$ and for the representative fixed flavors $j=g,u,s$. Additionally, the corresponding valence-number sum rule ratio defined in Eq.~\eqref{eq_numSumRule} also remains close to unity for the channels considered. The largest residual deviations are observed in the $u_vu$ and $u_v\bar u$ channels. Numerical-resolution and factorized-input tests show that these deviations decrease with improved resolution and are associated with the nontrivial flavor-dependent structure of the GS09 input, including the valence-number and quark-antiquark correlation
	contributions. Since these deviations remain at the percent level, we conclude that the unequal-scale evolution preserves both the momentum and valence-number constraints within the numerical accuracy of the present implementation.
	
	We then investigated several MRW-inspired prescriptions for generating the transverse momentum dependence of DPDFs. These include the double LO-MRW model in the perturbative region, denoted by DLO-MRW, the double virtuality ordered MRW model, DVO-MRW, its normalization-matched version, MDVO-MRW, and the double modified KMRW model, DMKMRW. The DVO-MRW approach provides a compact last step prescription that can be applied directly to the full DPDF, but it is not exactly normalization preserving. The MDVO-MRW model removes this integrated normalization mismatch by a multiplicative matching factor while keeping the virtuality ordered transverse momentum shape. In contrast, the DMKMRW model generates the transverse dependence through normalized external kernels and therefore reproduces the input collinear DPDF after integration over both transverse momenta.
	
	The comparison between factorized and non-factorized UDPDFs shows that the impact of longitudinal DPDF correlations depends strongly on the prescription used to generate the transverse momentum. In the DMKMRW model, the non-factorized to factorized ratio is independent of $k_t$, because the transverse kernels cancel in the ratio. In this case the effect of non-factorized DPDFs is purely longitudinal and appears as a channel and $x$-dependent correction factor. In the DLO-MRW, DVO-MRW and MDVO-MRW models, however, the DPDF is evaluated inside the last step convolution at parent momentum fractions $x_1/z_1$ and $x_2/z_2$ and at $k_t$-dependent scales. As a result, changing $k_t$ changes the region of the DPDF that is sampled, producing a $k_t$-dependent correlation effect.
	
	The flavor dependence is also important. The $uu$ channel shows a stronger suppression of the non-factorized contribution than the other channels, which is naturally connected with the valence-number sum rule. Once one valence $u$ quark is selected, the probability of finding another $u$ quark is constrained by the remaining valence content of the proton. In contrast, the $u\bar u$ channel can be enhanced relative to the factorized approximation because of the quark-antiquark correlation term present in the input DPDF at the initial scale. These results show that DPDF correlations cannot be represented by a universal multiplicative factor, and they depend on flavor, longitudinal momentum fraction, and transverse momentum.
	
	The model comparisons further indicate that the choice of UDPDF prescription is a relevant source of theoretical uncertainty. Although DMKMRW and MDVO-MRW are both normalization preserving, they lead to different transverse momentum shapes. The difference is especially visible at large transverse momentum, where the virtuality ordered construction is restricted by the condition $K^2<\mu^2$, while the MKMRW transverse kernel is normalized over the full transverse momentum range.
	
	Overall, our study indicates that both non-factorized DPDF correlations and the model dependence of the last step transverse momentum construction can be important in DPS phenomenology. In small $x$ and moderate $k_t$ regions, the factorized approximation may provide a reasonable first estimate for some channels. However, at larger $x$, near the double parton kinematic boundary, and in channels sensitive to valence or quark-antiquark correlations, the use of non-factorized UDPDFs becomes important. A full assessment of the phenomenological impact requires implementing these distributions in process level DPS calculations with realistic final state kinematics and experimental cuts. Such applications are left for future work
	\section*{Data Availability Statement}
	No Data are associated in the manuscript.

\bibliographystyle{apsrev4-2}

\end{document}